\documentclass[iop,revtex4]{emulateapj}
\usepackage{graphics,graphicx}
\usepackage{helvet}
\usepackage[utf8]{inputenc}
\usepackage[T1]{fontenc}
\usepackage{soul}
\usepackage{hyperref}
\usepackage{amsmath, amssymb, gensymb}
\usepackage{ulem}
\usepackage{natbib}
\usepackage{microtype}
\usepackage{multirow}
\usepackage{morefloats}
\usepackage{rotating}
\usepackage{color}
\usepackage{appendix}
\usepackage{xspace}
\def\Msol {$\hbox{M}_\odot$\xspace}

\def\kms {km\,s$^{-1}$\xspace}

\def\jybmkms {Jy\,beam$^{-1}$\,km\,s$^{-1}$\xspace}

\def\mjybm {mJy\,beam$^{-1}$\xspace}
\def\ujybm {$\mu$Jy\,beam$^{-1}$\xspace}

\def\jybmkms {Jy\,beam$^{-1}$\,km\,s$^{-1}$\xspace}

\def\etal {\textit{et al.}\xspace}

\def\co {CO\,($J$\,=\,2\,$\rightarrow\,$1)\xspace}
\def\cosixtofive {CO\,($J$\,=\,6\,$\rightarrow\,$5)\xspace}

\def\thirteenco {$^{13}$CO\,($J$\,=\,2\,$\rightarrow\,$1)\xspace}
\def\ceighteeno {C$^{18}$O\,($J$\,=\,2\,$\rightarrow\,$1)\xspace}

\def\ntwodp {N$_2$D$^+$\,($J$\,=\,3\,$\rightarrow\,$2)\xspace}

\newcommand{\bhr} {BHR~71\xspace}
\defcitealias{Tobin2019}{T19}
\newcommand{\tobin} {\citetalias{Tobin2019}\xspace}

\begin{document}

\slugcomment{Submitted to ApJ on 17 May 2019; accepted to ApJ on 15 Oct 2019}

\title{Understanding the origin of the magnetic field morphology in the \\ wide-binary protostellar system BHR~71}
\shorttitle{The magnetic field in BHR~71}

\author{Charles L. H. Hull\altaffilmark{1,2,11}}
\author{Valentin J. M. Le Gouellec\altaffilmark{3,4}}
\author{Josep M. Girart\altaffilmark{5,6}} 
\author{John J. Tobin\altaffilmark{7,8,9}}
\author{\\ Tyler L. Bourke\altaffilmark{10}}

\altaffiltext{1}{National Astronomical Observatory of Japan, NAOJ Chile, Alonso de C\'ordova 3788, Office 61B, 7630422, Vitacura, Santiago, Chile}
\altaffiltext{2}{Joint ALMA Observatory, Alonso de C\'ordova 3107, Vitacura, Santiago, Chile}
\altaffiltext{3}{European Southern Observatory, Alonso de C\'ordova 3107, Vitacura, Santiago, Chile}
\altaffiltext{4}{AIM, CEA, CNRS, Universit\'e Paris-Saclay, Universit\'e Paris Diderot, Sorbonne Paris Cit\'e, F-91191 Gif-sur-Yvette}
\altaffiltext{5}{Institut de Ci\`encies de l'Espai (ICE-CSIC), Campus UAB, Carrer de Can Magrans S/N, E-08193 Cerdanyola del Vall\`es, Catalonia}
\altaffiltext{6}{Institut d'Estudis Espacials de Catalunya, E-08030 Barcelona, Catalonia}
\altaffiltext{7}{National Radio Astronomy Observatory, 520 Edgemont Road, Charlottesville, Virginia 22903, USA}
\altaffiltext{8}{Homer L. Dodge Department of Physics and Astronomy, University of Oklahoma, 440 W. Brooks Street, Norman, OK 73019, USA}
\altaffiltext{9}{Leiden Observatory, Leiden University, P.O. Box 9513, NL-2300RA Leiden, The Netherlands}
\altaffiltext{10}{Square Kilometre Array Organization, Jodrell Bank Observatory, Lower Withington, Macclesfield, Cheshire SK11 9DL, UK}
\altaffiltext{11}{NAOJ Fellow}

\shortauthors{Hull \etal}
\email{chat.hull@nao.ac.jp}

\begin{abstract}
We present 1.3\,mm ALMA observations of polarized dust emission toward the wide-binary protostellar system \bhr IRS1 and IRS2.  IRS1 features what appears to be a natal, hourglass-shaped magnetic field.  In contrast, IRS2 exhibits a magnetic field that has been affected by its bipolar outflow.  Toward IRS2, the polarization is confined mainly to the outflow cavity walls.  Along the northern edge of the redshifted outflow cavity of IRS2, the polarized emission is sandwiched between the outflow and a filament of cold, dense gas traced by N$_2$D$^+$, toward which no dust polarization is detected.  This suggests that the origin of the enhanced polarization in IRS2 is the irradiation of the outflow cavity walls, which enables the alignment of dust grains with respect to the magnetic field---but only to a depth of $\sim$\,300\,au, beyond which the dust is cold and unpolarized.  However, in order to align grains deep enough in the cavity walls, and to produce the high polarization fraction seen in IRS2, the aligning photons are likely to be in the mid- to far-infrared range, which suggests a degree of grain growth beyond what is typically expected in very young, Class 0 sources.  Finally, toward IRS1 we see a narrow, linear feature with a high (10--20\%) polarization fraction and a well ordered magnetic field that is not associated with the bipolar outflow cavity.  We speculate that this feature may be a magnetized accretion streamer; however, this has yet to be confirmed by kinematic observations of dense-gas tracers.
\\ 
\end{abstract}


\section{Introduction}
\label{sec:intro}

Early theories of magnetized star-formation suggested that the formation of stars within molecular clouds should be regulated by a strong magnetic field \citep{Mestel1956, Shu1987, McKee1993, McKee2007}.  In such a scenario, if one observed at small enough spatial scales ($\lesssim$\,1000\,au), it was thought that one should see an ``hourglass'' morphology in the magnetic field lines, where a strong, poloidal magnetic field is pinched by the gravitational infall very near the central source \citep{Fiedler1993, Galli1993a, Galli1993b, Allen2003}.  And in fact, several examples of the fabled hourglass were seen in some of the first sources whose magnetic fields were observed at high resolution by the Berkeley-Illinois-Maryland Association (BIMA) millimeter array, the Submillimeter Array (SMA) and the Combined Array for Research in Millimeter-wave Astronomy (CARMA), including the bright, deeply embedded Class 0 protostellar sources NGC 1333-IRAS 4A \citep{Girart1999, Girart2006}, IRAS 16293A \citep{Rao2009}, and L1157 \citep{Stephens2013}.  These sources exhibit other hallmarks of strong-field star formation, including powerful outflows (whose generation is intimately connected to the magnetic field; \citealt{Frank2014}) and high (inferred) magnetic field strengths, on the order of a few milli-Gauss.  

Furthermore, the sources are not extremely fragmented, either being a single source (L1157, \citealt{Tobin2013c}), a binary (IRAS 4A, \citealt{Looney2000, Girart2006}), or a triple system (IRAS 16293, \citealt{Wootten1989, Chandler2005}).  This is consistent with both theoretical work that found that a strong magnetic field strongly limits fragmentation on $\sim$\,1000--10,000\,au scales \citep[e.g.,][]{Hennebelle2011}, as well as with a significant number of studies showing that a strong magnetic field could impede the formation of large disks, thus potentially reducing the frequency with which close-multiple systems form via disk fragmentation \citep{Kratter2010, Tobin2016b} at the scales of a few $\times$ 100\,au.  See \citet{WursterLi2018} for a recent review of the effect of the magnetic field on disk formation.  However, other studies have found that a strong magnetic field can actually increase fragmentation under certain circumstances \citep{Boss2000, Offner2016, Offner2017}.  See \citet{KrumholzFederrath2019} for a recent review discussing this topic in more detail.

After more than a decade of high-resolution observations of magnetic fields in forming stars with BIMA, the SMA, CARMA, and now the Atacama Large Millimeter/submillimeter Array (ALMA), it has become clear that hourglass-shaped fields appear to be the exception rather than the rule \citep{HullZhang2019}.  When observing at scales $\gtrsim$\,100\,au, these instruments probe polarization of dust grains that have been aligned with their minor axes parallel to the ambient magnetic field via the phenomenon of ``Radiative Alignment Torques'' \citep[RATs;][]{Draine1996, LazarianHoang2007a, Hoang2009, Andersson2015}; this makes the polarized millimeter and submillimeter (hereafter, ``(sub)millimeter'') emission from the dust grains an excellent tracer of the magnetic field.  Surveys of dust polarization toward both low- and high-mass young stellar objects (YSOs) found that outflows and magnetic fields are randomly aligned, calling into question the ability of magnetic fields to regulate star formation on small scales \citep{Hull2013, Hull2014, Zhang2014, HullZhang2019}.  Furthermore, recent ALMA observations have revealed a source with chaotic magnetic fields whose structure is most likely dominated by turbulence and infall \citep{Hull2017a}, and other sources whose magnetic fields have been affected (and possibly shaped) by outflows \citep{Hull2017b, Maury2018, LeGouellec2019a}.    

The fact that the \bhr binary system \citep{Bourke2001} is among the brightest Class 0 protostars known---similar to IRAS 4A, IRAS 16293, and L1157---might suggest that we should see an hourglass around one or both members of the binary.  This is because strong, poloidal fields could plausibly help funnel infalling material onto the central sources more efficiently, thus increasing their brightness.  However, in contrast to this well ordered, quiescent formation scenario, recent evidence points to a turbulent origin for \bhr, which comprises two binary components---IRS1 and IRS2---that are separated by $\sim$\,15$\arcsec$, or $\sim$\,3000\,au at a distance of 200\,pc \citep{Seidensticker1989, Bourke1997}.  And indeed, in general, wide binaries (with separations $\gtrsim$\,1000\,au) like \bhr are suspected to be the result of turbulent fragmentation \citep[e.g.,][]{Pineda2015, Lee2016, Tobin2016a}.  

The first piece of evidence pointing to a turbulent origin for \bhr is the misalignment of the outflows emanating from IRS1 and IRS2 \citep{Bourke2001, Parise2006}.  Previous studies have identified misaligned outflows \citep{Tobin2015a, Lee2016} and misaligned disks \citep{JensenAkeson2014, JELee2017} in wide binary/multiple sources; these cases are consistent with simulations of multiple sources forming in turbulent environments \citep{Offner2016}.  Second, there is tentative evidence that the envelopes around the two binary components are counter-rotating, as seen in \ceighteeno \citep[][hereafter \citetalias{Tobin2019}]{Tobin2019}; these misaligned velocity gradients of the dense envelope material around IRS1 and IRS2 are also consistent with the turbulent formation of wide binaries.  This is in contrast to close binaries (with separations $\lesssim$\,200\,au), which are most likely the result of disk fragmentation, and whose components tend to have consistent angular momentum orientations that are roughly perpendicular to the binary/multiple's orbital plane \citep{Tobin2018}.  Finally, the outflows from IRS1 and IRS2 are misaligned with the filamentary dust and ammonia structure in which they are embedded (\tobin), consistent with a  turbulent origin for the sources, which have not obviously inherited their angular momentum orientation from their natal filament.  This is consistent with work in the Perseus molecular cloud by \citet{Stephens2017a} showing that outflows and filamentary structure are randomly oriented with respect to one another. 

Our understanding of the role of the magnetic field in the formation of binary/multiple systems is in its infancy.  One of the first studies to touch on this topic is the recent work by \citet{Galametz2018}, who found tentative evidence in their SMA data that there are large misalignments between the outflows and the magnetic field orientations in protostellar cores with higher rotational energies.  They observed a $\sim$\,90$^\circ$ misalignment in some objects, which could be attributed to rotational winding of the magnetic field lines. Additionally, they found that several of the objects in this subsample of sources (with magnetic fields and outflows oriented perpendicular to one another) happen to be wide multiple sources and/or have large disks, whereas the sources in their sample with aligned outflows and magnetic fields tend to be single objects with small (or unresolved) disks.  This is consistent with the tentative trend seen in the VLA observations of Class 0 and I protostellar cores by \citet{SeguraCox2018}, where sources with larger disks tend to have misaligned magnetic fields and outflows.  These results hint at a relationship between multiplicity and the magnetic field that can be further revealed by targeted ALMA studies such as the one we present here, as well as future surveys of dust polarization toward sources whose multiplicity has already been determined, such as those observed as part of the VLA Nascent Disk and Multiplicity (VANDAM) survey \citep{Tobin2016a} using the Karl G. Jansky Very Large Array (VLA). 

Here we present full-polarization observations of \bhr, with the goal of understanding the origin of the magnetic field morphologies toward the two components of this iconic wide-binary system.  There have been many observations of \bhr over the last two decades \citep[e.g.,][]{Bourke1997, Garay1998, Bourke2001, Parise2006, Chen2008}; these include recent ALMA observations of spectral lines by \citetalias{Tobin2019} (see Section \ref{sec:tobin_obs}) with spatial resolutions similar to those of the polarization observations we present here.  As we describe below, our observations reveal an unexpected combination of what appears to be a natal hourglass magnetic field around \bhr IRS1, and a magnetic field that clearly has been affected by the outflow around IRS2.

We discuss our ALMA dust polarization observations toward \bhr in \S\,\ref{sec:obs} and \S\,\ref{sec:res}.  In \S\,\ref{sec:dis} we compare our polarization observations with several ALMA spectral-line observations of outflows and dense-gas tracers published by \tobin, and we discuss the possible origins of the polarized emission in the two sources.  We offer our conclusions and potential paths forward in \S\,\ref{sec:con}.

\subsection{Previous observations by \citet[][or \tobin]{Tobin2019}}
\label{sec:tobin_obs}

\tobin discussed the formation conditions of \bhr at length, using both new and archival data to analyze the kinematics and continuum properties across a wide range of spatial scales ranging from $\sim$\,0.5\,pc\,--\,80\,au.  
They presented J-band near-infrared (NIR) observations from the Infared Side-Port Imager (ISPI) instrument on the Blanco 4\,m telescope at Cerro Tololo; 
H and Ks-band NIR data from the Persson's Auxilary Nasmyth Infrared Camera (PANIC) instrument on the Magellan Baade 6.5\,m telescope at the Las Campanas Observatory; 
mid-infrared \textit{Spitzer} IRAC, MIPS, and IRS observations; 
far-infrared \textit{Herschel} PACS and SPIRE observations;
millimeter-wave ALMA observations of \co, \thirteenco, \ceighteeno, and \ntwodp; 
centimeter-wave NH$_3$\,(1,1) observations from the Parkes radio telescope; 
and NH$_3$\,(1,1),\,(2,2),\,(3,3) observations from the Australian Telescope Compact Array (ATCA).

While the Parkes and ATCA observations showed a clear (albeit small, $\sim$\,1\,km\,s$^{-1}$\,pc$^{-1}$) velocity gradient across the core in which both IRS1 and IRS2 are embedded, analysis by \tobin of the suite of higher-resolution spectral-line observations showed much more complex structure, including tentative signatures in the ALMA \ceighteeno emission at <\,1000\,au scales that the envelopes around IRS1 and IRS2 may be rotating in opposite directions.  These observations led the authors to conclude that the \bhr binary most likely formed from turbulent fragmentation rather than from rotational/disk fragmentation.

Assuming dust temperatures of 34\,K for IRS1 and 20\,K for IRS2, \tobin calculated dust\,+\,gas masses of 0.59 and 0.11\,\Msol for the two sources, respectively.  The available multi-wavelength observations also enabled \tobin to confirm that both sources are protostars in the youngest (Class 0) stage of protostellar evolution, and to estimate their bolometric luminosities: 14.7\,$L_\odot$ for IRS1 and 1.7\,$L_\odot$ for IRS2.

Estimates by \tobin of the inclination of the outflows from IRS1 and IRS2 with respect to the plane of the sky yielded values of $\sim$\,50--60$\degree$ for both sources (where 90$\degree$ means the outflow is fully in the plane of the sky).  This is consistent with an estimate from \citet{YLYang2017}, who estimated a 50$\degree$ inclination angle for IRS1.  The full opening angles (i.e., the angle between the two edges of the outflow) for IRS1 and IRS2 are 55$\degree$ and 47$\degree$, respectively.

\section{Observations}
\label{sec:obs}

We present ALMA observations of dust polarization at 1.3\,mm (Band 6) toward \bhr taken on 2018 May 03.  The observations included 43 antennas.  The precipitable water vapor (PWV) ranged from 0.9 to 1.7\,mm during the observations.  The pointing center was ($\alpha_\textrm{J2000}$\;=\;12:01:36.514, $\delta_\textrm{J2000}$\;=\;--65:08:49.31) for IRS1 and ($\alpha_\textrm{J2000}$ = 12:01:34.042, $\delta_\textrm{J2000}$ = --65:08:47.870) for IRS2.  The observations included $\sim$\,100\,min of on-source time, taken during the LST range of $\sim$\,11:30--16:15 (i.e., $\sim$\,00:30 before transit to $\sim$\,4:15 after transit).  With a Briggs weighting parameter of \texttt{robust}\,=\,2.0 (see further discussion later in this section), the maps have a synthesized beam (or resolution element) of 1$\farcs$10\,$\times$\,0$\farcs$89, 
equivalent to a linear resolution of $\sim$\,200\,au at a distance of 200\,pc.  
The baselines in the C-2 antenna configuration range from 15--500\,m.  The maximum spatial scale at which emission can be recovered by ALMA is approximately 10$\farcs$6.

The polarization data include 8.0\,GHz of wide-band dust continuum ranging in frequency from $\sim$\,223--227\,GHz and $\sim$\,239--243\,GHz, with a mean frequency of 233\,GHz (1.3\,mm).  Each 2\,GHz spectral window (with 1.875\,GHz of usable bandwidth) was divided into 64 channels with widths of 31.25\,MHz.    The flux and bandpass calibrator was J1107-4449; the polarization calibrator was J1256-0547; and the phase calibrator was J1206-6138.  These flux/bandpass and phase calibrators were chosen automatically by querying the ALMA source catalog when the project was executed.  The polarization calibrator was chosen by hand due to its high polarization fraction. The observatory's flux monitoring program has determined that at Band 6, ALMA's systematic flux-calibration accuracy is $\sim$\,10\%.  See \citet{Nagai2016} for a detailed discussion of the ALMA polarization system.  The systematic uncertainty in on-axis linear polarization observations with ALMA is 0.03\% (corresponding to a minimum detectible polarization of 0.1\%).  

We produced the dust continuum images using the Common Astronomy Software Applications (CASA, \citealt{McMullin2007}) tasks \texttt{TCLEAN} and \texttt{CLEAN}. For all of the figures below that show both IRS1 and IRS2 in the same image, we constructed a two-point mosaic of the sources using \texttt{TCLEAN}, centered at ($\alpha_\textrm{J2000}$\;=\;12:01:35.202, $\delta_\textrm{J2000}$\;=\;--65:08:48.64), half way between the two sources.  While we mainly analyze the lower ($\sim$\,1\farcs0) resolution, \texttt{robust}\,=\,2.0 images, we also made higher resolution images Briggs weighting parameters of \texttt{robust}\,=\,0.5 and --0.5, yielding approximate resolutions of 0\farcs75 and 0\farcs53, respectively, in order to highlight higher resolution features toward IRS1.  

We also made images of IRS1 and IRS2 individually using \texttt{CLEAN}.  Making non-mosaicked images was necessary to produce high fidelity images of the Stokes $I$ emission from IRS2, which is much fainter than IRS1.  When IRS1 was observed in the center of the field of view, it was so bright that our ability to image faint Stokes $I$ emission was limited by dynamic range.  As a result, in the mosaicked images the quality of the Stokes $I$ image of IRS2 was degraded by the presence of IRS1 in the mosaic.  However, when IRS2 was imaged alone, IRS1 was significantly removed from the phase center (and was thus fainter, due to the primary beam response of the ALMA antennas), which mitigated the dynamic range effects and allowed us to produce higher quality images of low-level Stokes $I$ dust emission surrounding IRS2.

All images were improved by performing four rounds of phase-only self-calibration, where the total intensity (Stokes $I$) image was used as a model.  The shortest interval for determining the gain solutions was 12\,s.  See \citet{Brogan2018} for a detailed discussion of and best practices for self calibration.  The Stokes $I$, $Q$, and $U$ maps (where the $Q$ and $U$ maps show the polarized emission) were independently cleaned with an appropriate number of \texttt{CLEAN} or \texttt{TCLEAN} iterations after the final round of self-calibration.  The rms noise level in the dynamic-range-limited Stokes $I$ dust maps ranges from $\sigma_I \approx 300$\,\ujybm{} (\texttt{robust}\,=\,2.0; mosaic) to 250\,\ujybm{} (\texttt{robust}\,=\,0.5, --0.5; mosaics) to 150\,\ujybm{} (\texttt{robust}\,=\,2.0; single pointing toward IRS2).  The rms noise level $\sigma_P$ in the maps of polarized intensity $P$ (see Equation \ref{eqn:P} below), which are not dynamic-range limited, ranges from 25\,\ujybm (\texttt{robust}\,=\,2.0,\,0.5) to 40\,\ujybm (\texttt{robust}\,=\,--0.5).  The noise level in the $P$ maps increases as the \texttt{robust} parameter decreases because weighting the $uv$-data to produce higher resolution images tends to increase the noise level \citep{TMS}. 

The quantities we derive from the polarization maps include the polarized intensity $P$, the linear polarization fraction $P_\textrm{frac}$, and the polarization position angle $\chi$:

\begin{align}
P &= \sqrt{Q^2 + U^2} \label{eqn:P} \\
P_\textrm{frac} &= \frac{P}{I} \\
\chi &= \frac{1}{2} \arctan{\left(\frac{U}{Q}\right)}\, .
\end{align}

\noindent
$\chi$ is defined to have a position angle of 0$\degree$ when oriented N--S, and increases to the east \citep{IAU1974}.  Note that $P$ has a positive bias because it is always a positive quantity, whereas the Stokes parameters $Q$ and $U$ (from which $P$ is derived) can be positive or negative.  This bias is particularly significant in low-signal-to-noise (<\,5\,$\sigma$) measurements.  We debias the polarized intensity map as described in \citet{Vaillancourt2006} and \citet{Hull2015b}, although we note that it has only a very minor effect on our results. 

As has been the case with several polarization results from ALMA, we detect a marginal circularly-polarized signal in the Stokes $V$ map; however, the circular polarization fraction in the on-axis (single-pointing, non-mosaicked) observations of IRS1 is $\sim$\,0.08\% of the total intensity, nearly an order of magnitude smaller than the current 0.6\% systematic uncertainty in ALMA circular polarization observations, and thus is likely to be spurious.  Furthermore, the behavior of the Stokes $V$ signal when both IRS1 and IRS2 are observed off-axis is consistent with the known beam-squint profile \citep{Chu1973, Adatia1975, Rudge1978} of the ALMA 12\,m antennas at Band 6.

\section{Results}
\label{sec:res}

\begin{figure}[hbt!]
\begin{center}
\vspace{0.5em}
\includegraphics[width=0.47\textwidth, clip, trim=0cm 2.36cm 0cm 0cm]{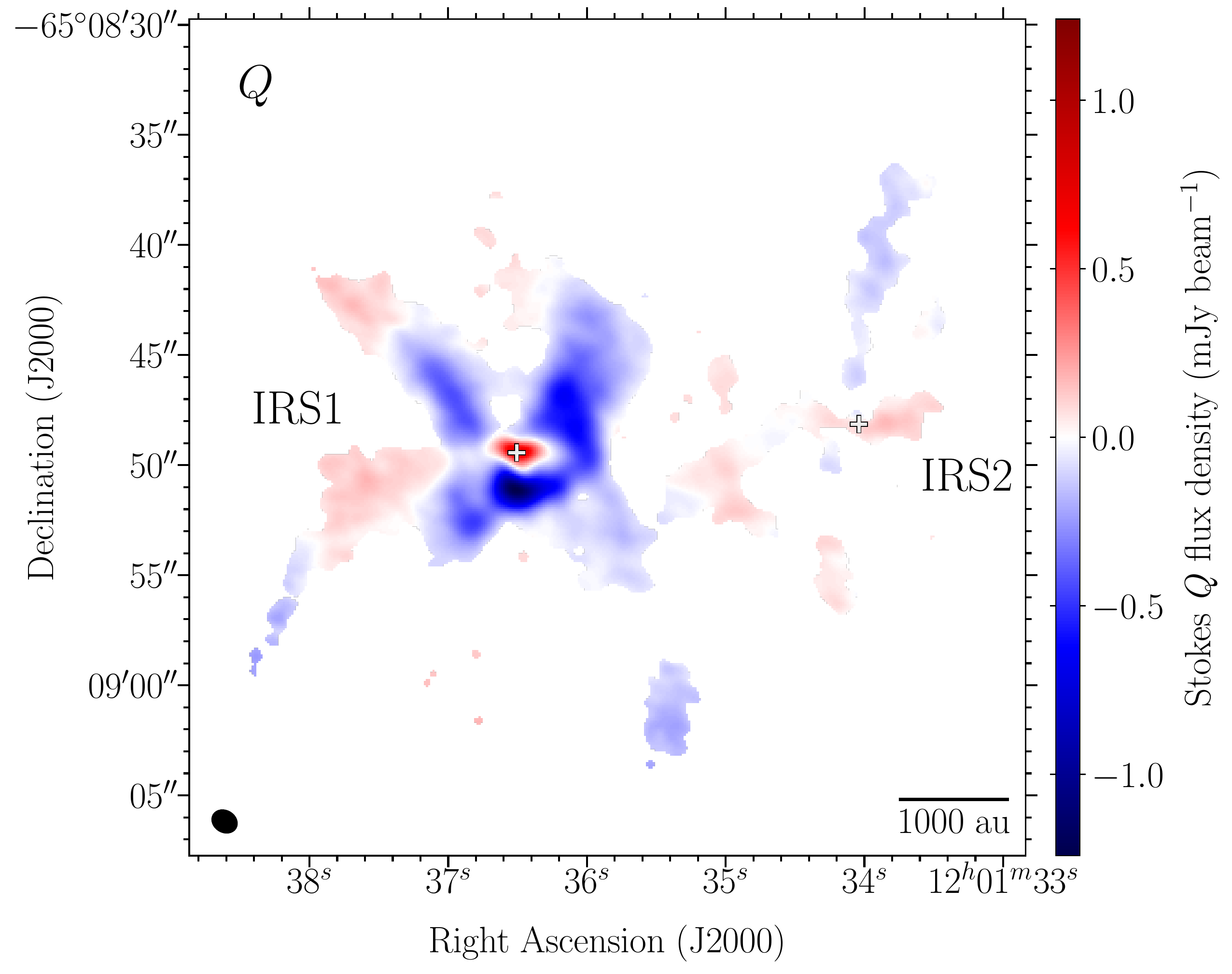}
\includegraphics[width=0.47\textwidth, clip, trim=0cm 2.36cm 0cm 0.25cm]{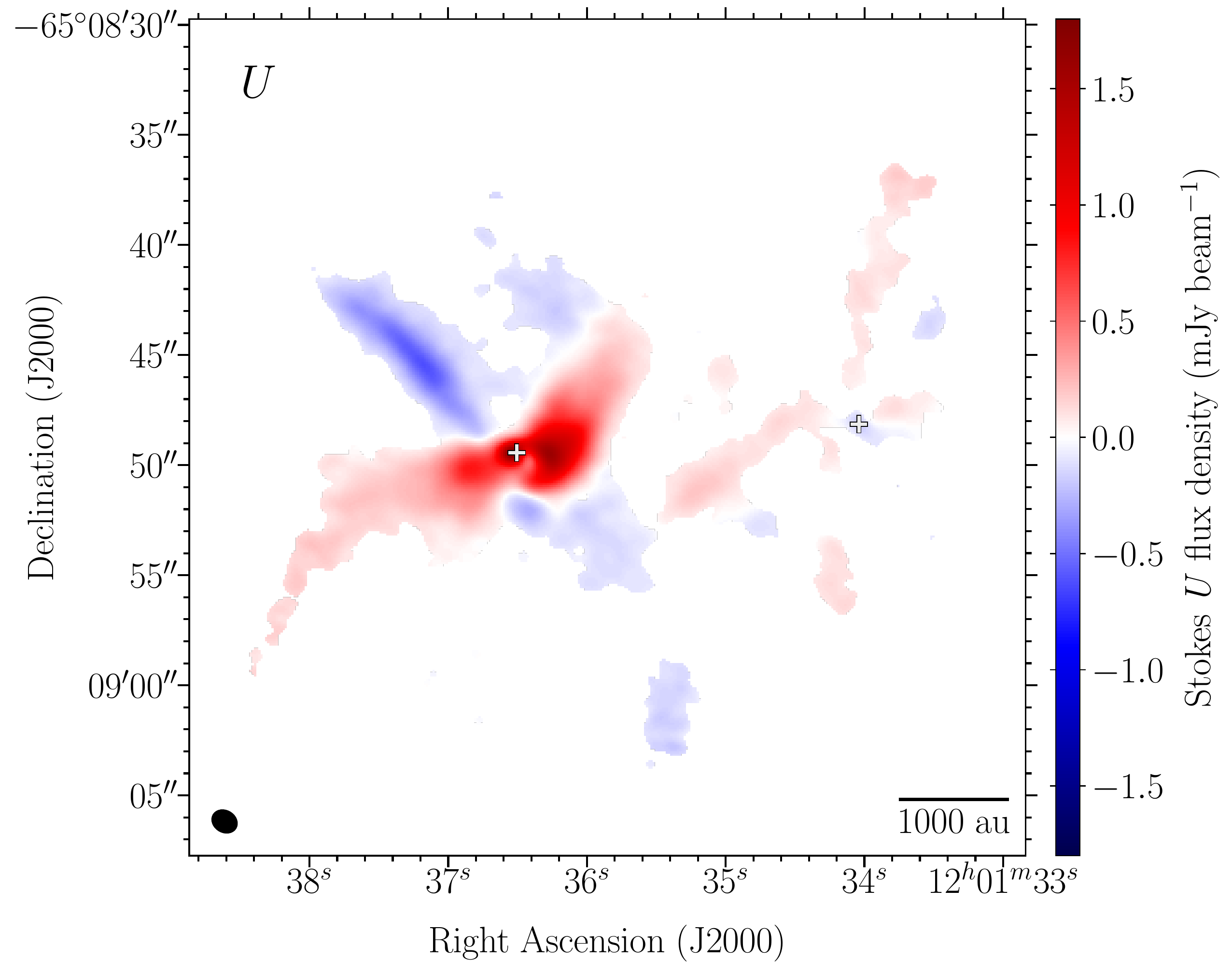}
\includegraphics[width=0.45\textwidth, clip, trim=0.73cm 0cm 2.58cm 0.33cm]{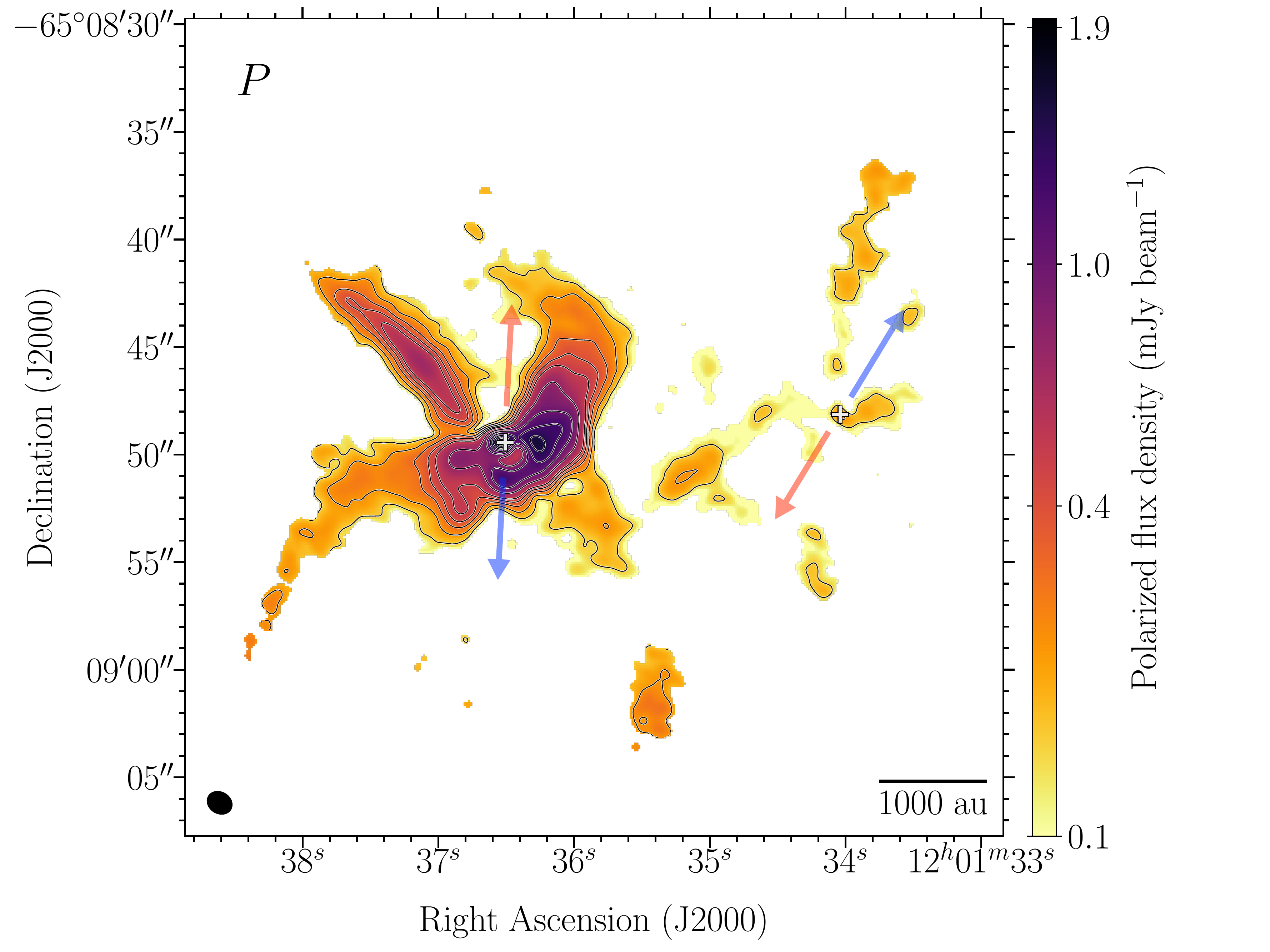}
\end{center}
\vspace{-1em}
\caption{\footnotesize
Maps of Stokes $Q$ (top), $U$ (middle), and polarized intensity $P$ (bottom) toward \bhr.  The maximum and the minimum of the $Q$ and $U$ color scales are symmetric around zero, and the scale ranges are set by $|Q|_\textrm{max} = 1.24$\,\mjybm and $|U|_\textrm{max} = 1.80$\,\mjybm.  The peak value of $P$ is 1.94\,\mjybm, and the rms noise level $\sigma_P = 25$\,\ujybm.  The contour levels are 5,\,8,\,12,\,16,\,20,\,30,\,40,\,50,\,60,\,70 $\times$ $\sigma_P$.  The black ellipses in the lower-left corners of all panels represent the ALMA synthesized beam (resolution element), which measures 1$\farcs$10\,$\times$\,0$\farcs$89, equivalent to a linear resolution of $\sim$\,200\,au at a distance of 200\,pc.  Crosses indicate the continuum peaks of IRS1 and IRS2.  Blue and red arrows in the bottom panel indicate the orientations of the blue- and redshifted lobes of the bipolar outflows from IRS1 and IRS2. 
\textit{The ALMA data used to make this figure are available in the online version of this publication.}
\vspace{0em} 
}
\label{fig:QU_POLI}
\end{figure}

\begin{figure*}[hbt!]
\begin{center}
\vspace{1em}
\includegraphics[width=0.505\textwidth, clip, trim=0cm -0.18cm 4.7cm 0cm]{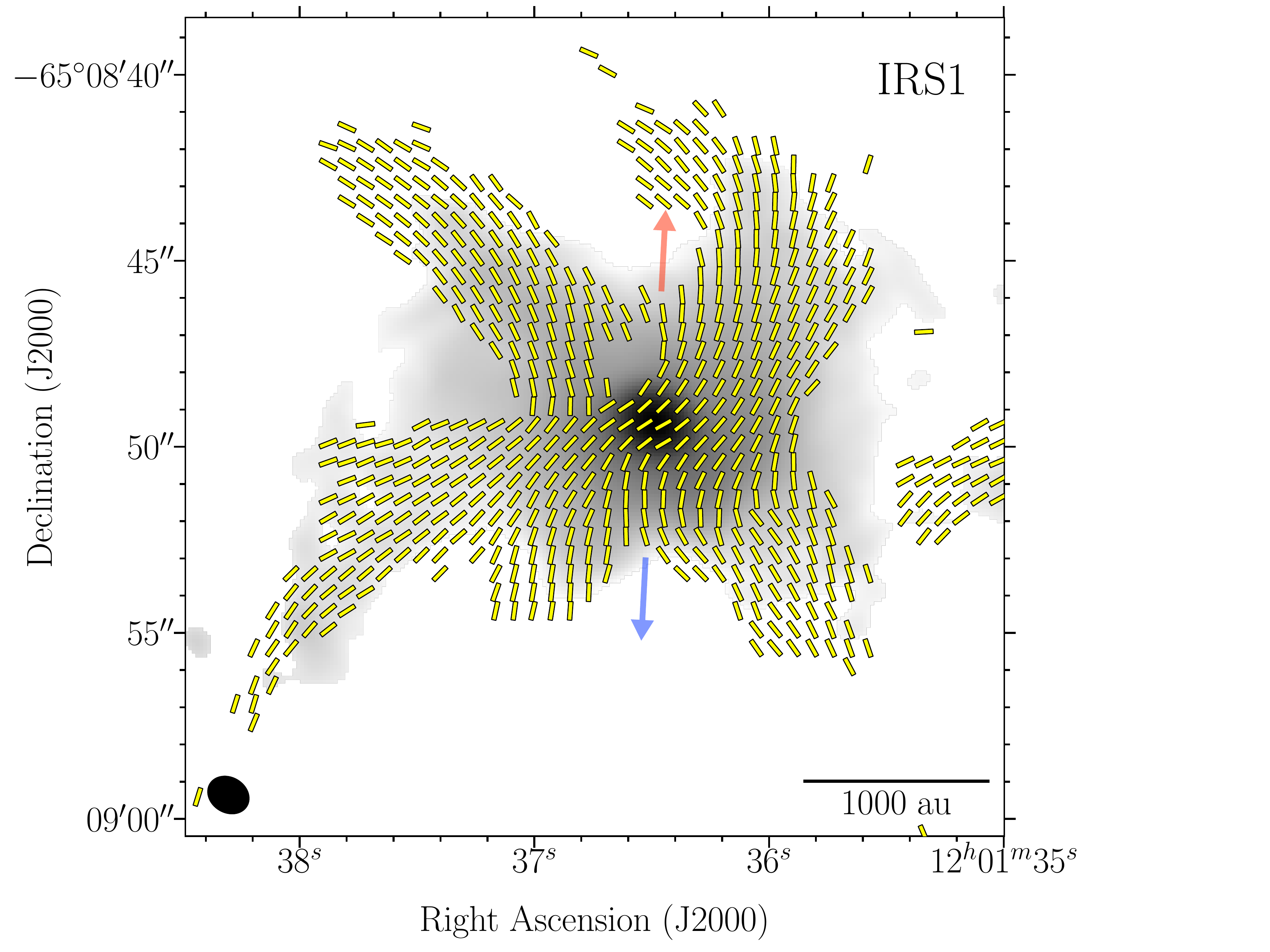}
\includegraphics[width=0.48\textwidth, clip, trim=3.71cm 0cm 0cm 0cm]{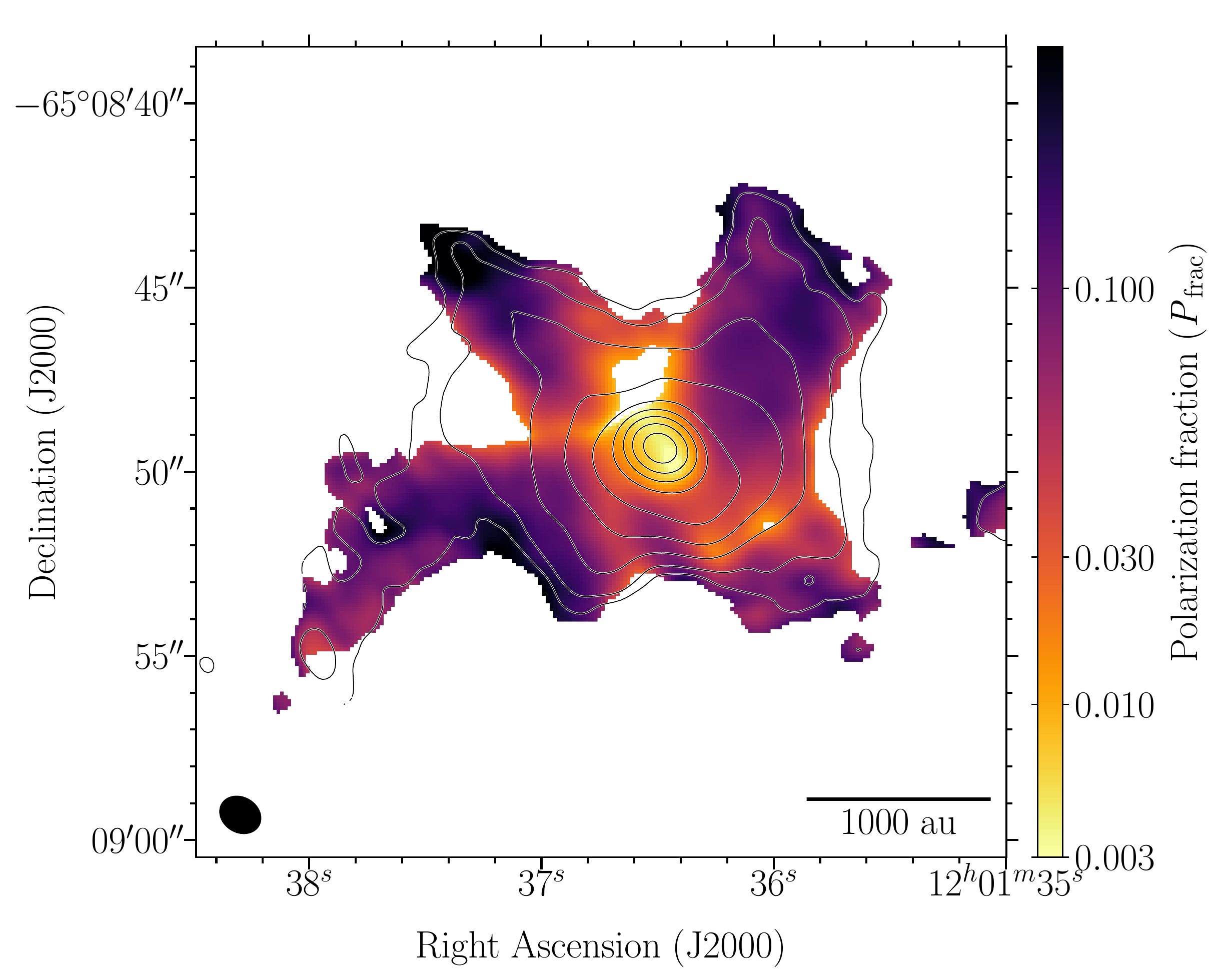}
\includegraphics[width=0.498\textwidth, clip, trim=0cm -0.18cm 5.5cm 0cm]{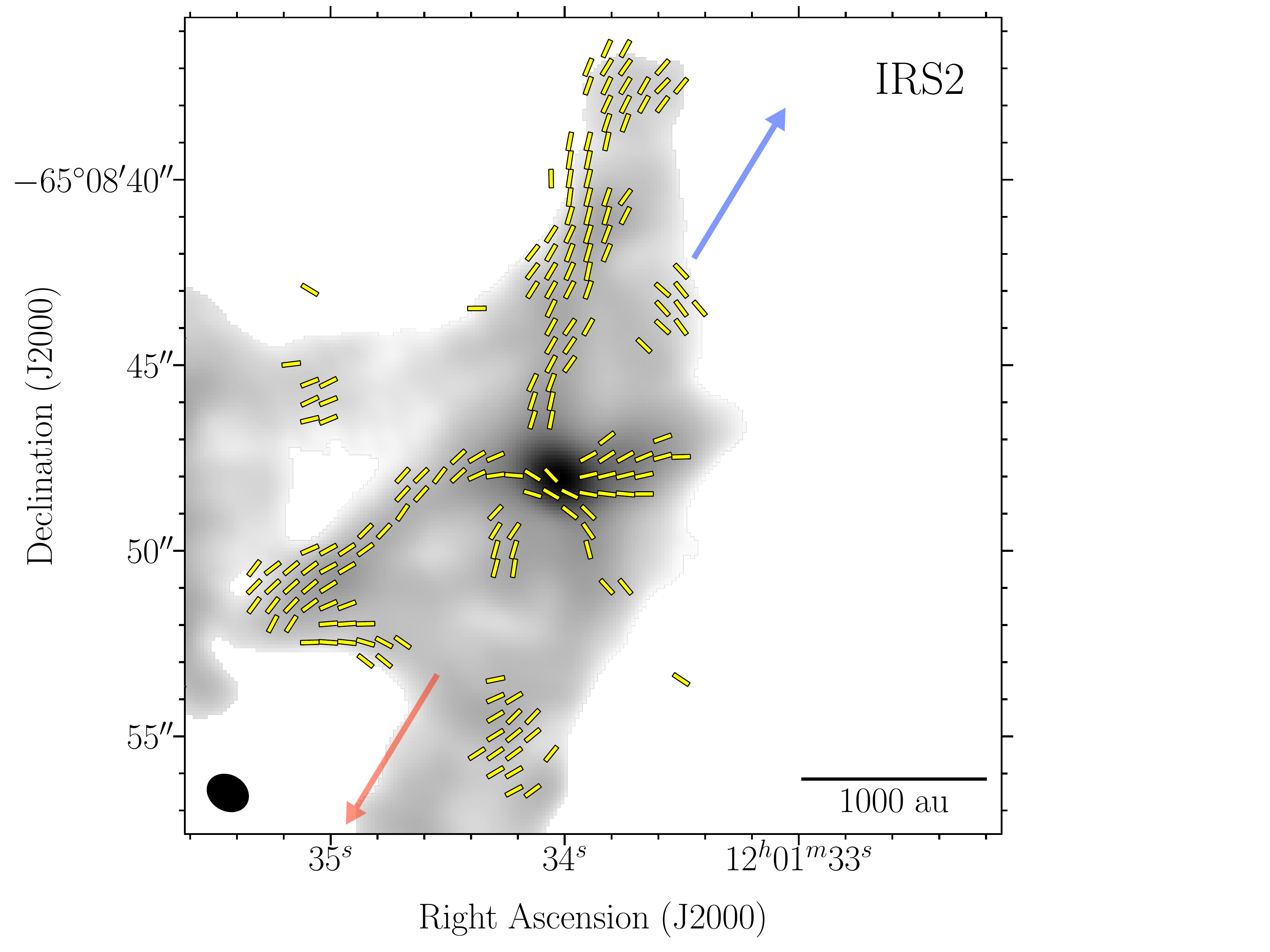}
\includegraphics[width=0.481\textwidth, clip, trim=3.71cm 0cm .2cm 0cm]{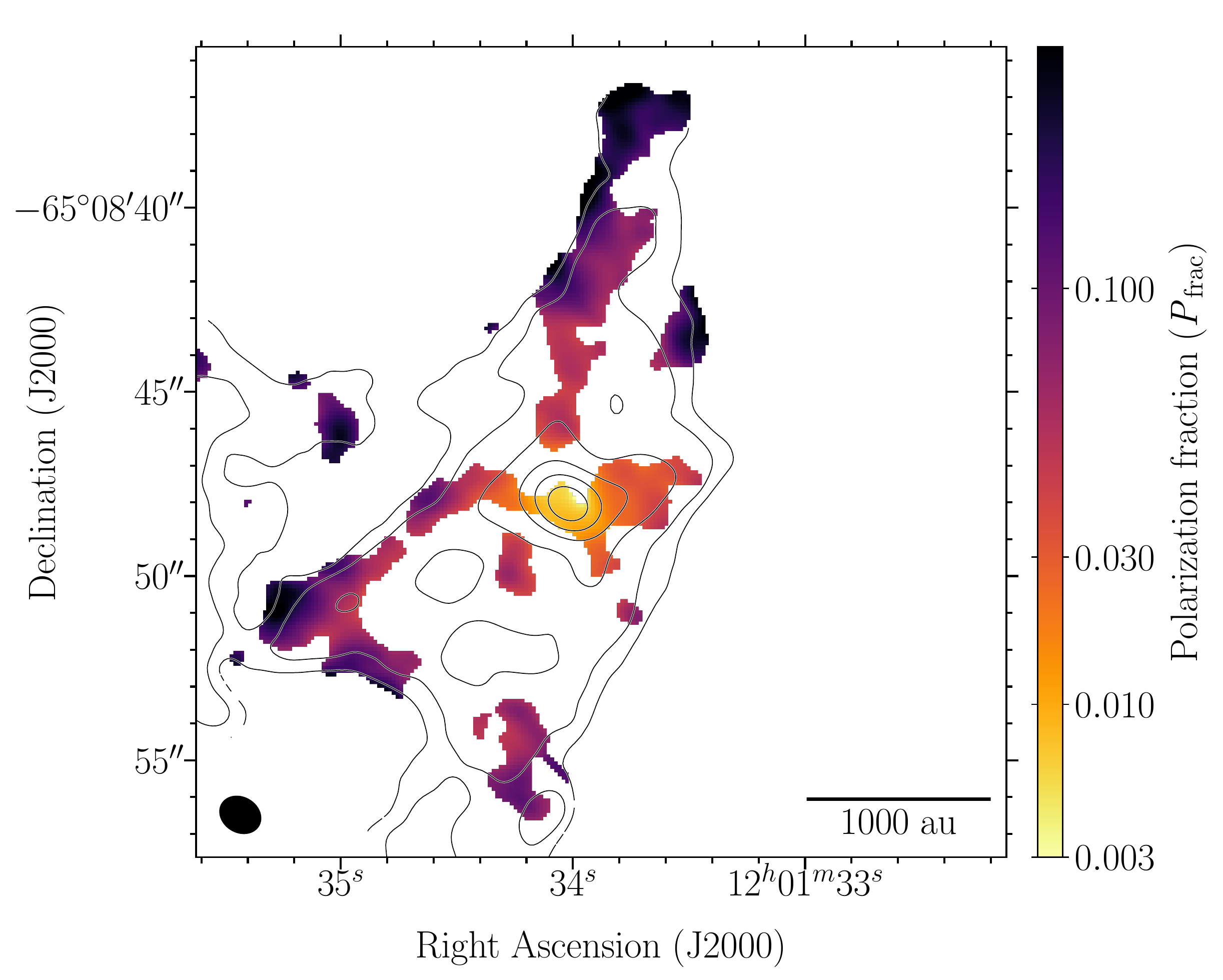}
\end{center}
\vspace{-1em}
\caption{\footnotesize
Magnetic field morphology (left) and linear polarization fraction (right) toward \bhr IRS1 and IRS2. \textit{Left column:} grayscale is the total intensity (Stokes $I$) thermal dust emission, plotted beginning at 3\,$\sigma_I$, where $\sigma_I$ is the rms noise in the Stokes $I$ map.  For IRS1, $\sigma_I = 300$\,\ujybm; for IRS2, $\sigma_I = 150$\,\ujybm.  Line segments are the inferred magnetic field, rotated by 90$\degree$ from the polarization orientation, and are plotted starting at 3\,$\sigma_P$, where $\sigma_P$ is the rms noise in the polarized intensity $P$ map.  For both IRS1 and IRS2, $\sigma_P = 25$\,\ujybm.  The line segments are all the same length, and do not represent any other quantity.  Blue and red arrows indicate the orientations of the blue- and redshifted lobes of the bipolar outflows from IRS1 and IRS2.  \textit{Right column:} polarization fraction ($P_\textrm{frac}$) is in color scale, and is plotted where both Stokes $I$ > 3\,$\sigma_I$ and $P$ > 3\,$\sigma_P$.  Contours are the Stokes $I$ emission, plotted at 5,\,8,\,16,\,32,\,64,\,128,\,256,\,512,\,1024 $\times$ the $\sigma_I$ value in the respective map.  The black ellipses in the lower-left corners of all panels represent the synthesized beam of the dust emission, which measures 1$\farcs$10\,$\times$\,0$\farcs$89.  
\textit{The ALMA data used to make this figure are available in the online version of this publication.}
\vspace{1em} 
}
\label{fig:pol_pfrac}
\end{figure*}

We show the results of our full-polarization, 1.3\,mm dust continuum observations toward the \bhr binary in Figures~\ref{fig:QU_POLI} and \ref{fig:pol_pfrac}.  Later, we overlay these dust polarization results on images of the dense-gas tracers \ceighteeno and \ntwodp (Figure~\ref{fig:c18o_n2d+}) and on \co outflow emission (Figures~\ref{fig:co} and \ref{fig:irs2_zoom}).  We also show dust polarization maps at multiple resolutions (Figure~\ref{fig:3res}).

In Figure~\ref{fig:QU_POLI} we show the Stokes $Q$, $U$, and polarized intensity $P$ maps.  By comparing by eye with the synthetic models of $Q$ and $U$ for an hourglass-shaped magnetic field morphology shown in \citet[][Figure~4]{Frau2011}, we can constrain the inclination of the magnetic field in \bhr IRS1 with respect to the the plane of the sky.  The centers of the $Q$ and $U$ maps toward IRS1 look quite similar to the \citeauthor{Frau2011} results for $\omega = 30\degree$, which is equivalent to a $\sim$\,60$\degree$ inclination of the source with respect to the plane of the sky.  This is consistent with the inclination estimates by both \tobin and \citet{YLYang2017}.

In Figure~\ref{fig:pol_pfrac} we overlay the inferred magnetic field orientations (produced by rotating the polarization orientations by 90$\degree$) on the total intensity (Stokes $I$) dust continuum map.  When one studies only the maps of dust polarization, the magnetic field morphologies of both sources are consistent with the hourglass shape discussed in Section \ref{sec:intro}.  However, when the magnetic field maps are analyzed alongside spectral-line observations, it becomes clear that while the magnetic field toward IRS1 looks more like a ``traditional'' hourglass, in IRS2 the majority of the polarization detections are consistent with a magnetic field lying along the cavity walls of the bipolar outflow, traced by \co emission (see Figures~\ref{fig:co} and \ref{fig:irs2_zoom}).  While the extended hourglass structure toward IRS1 has a symmetry axis that is well aligned with the CO outflow, the majority of the polarized emission is spatially extended far beyond the outflow cavity, and thus, in contrast to IRS2, the magnetic field toward IRS1 was most likely not shaped by the outflow.  See Section \ref{sec:hourglass_vs_outflow} for further discussion.

In Figure~\ref{fig:pol_pfrac} we also show maps of the polarization fraction $P_\textrm{frac}$ toward IRS1 and IRS2.  These maps exhibit the typical features of polarization maps of protostellar cores, including a ``polarization hole,'' or a drop in $P_\textrm{frac}$ to values $\lesssim$\,1\% toward the Stokes $I$ intensity peak \citep[see][and references therein]{Hull2014}.  In spite of having polarization holes toward the very centers of the sources, the maps of IRS1 and IRS2 show polarization fraction levels >\,10\% across much of both sources.  These high levels of polarization can be reproduced by the models of grain alignment via RATs, and have been seen in interferometric observations of polarization toward both low- and high-mass star-forming regions \citep[e.g.,][]{Stephens2013, Hull2014, Cortes2016, Kwon2019, Cortes2019}.  However, our ability to interpret $P_\textrm{frac}$ is limited by the fact that the Stokes $I$ image is much more strongly dynamic-range limited than the $P$ image, resulting in a $P$ map that extends beyond the limits of the $I$ map.  This has been seen in other high-sensitivity ALMA polarization maps \citep[e.g.,][]{Hull2017b, Kwon2019, LeGouellec2019a}.

We can take advantage of the excellent $uv$-coverage of ALMA to make images at different resolutions, varying the resolution by up to a factor of 2 by using different \texttt{robust} weighting parameters in the imaging process.  When we make the polarization map of \bhr with three resolutions (1\farcs0, 0\farcs75, and 0\farcs53), we clearly see an exceptionally sharp filamentary/stream-like structure extending to the NE of IRS1 (see Figure~\ref{fig:3res}).  These sharp structures, which are generally more apparent in $P$ images than in $I$ images, have been seen in other high-resolution, high-sensitivity ALMA polarization observations of protostellar cores \citep{Hull2017a, Hull2017b, Maury2018, Sadavoy2018c, Takahashi2019, LeGouellec2019a}, and are discussed further in Section \ref{sec:pol_origins}.

\section{Discussion}
\label{sec:dis}

We begin this discussion by focusing on several plots, including the dust continuum, inferred magnetic field, and polarization fraction toward both binary components of \bhr (Figure~\ref{fig:pol_pfrac}); the magnetic field overlaid on maps of the dense-gas tracers \ceighteeno and \ntwodp (Figure~\ref{fig:c18o_n2d+}); and the magnetic field overlaid on the bipolar outflows from IRS1 and IRS2 traced by \co (Figures~\ref{fig:co} and \ref{fig:irs2_zoom}).  

As first mentioned in Section \ref{sec:res} and discussed further in Sections \ref{sec:dense_tracers} and \ref{sec:hourglass_vs_outflow}, when the polarization maps are analyzed alongside maps of spectral-line emission, it becomes clear that the magnetic field morphologies in IRS1 and IRS2 have different origins.  The brighter source IRS1 has a magnetic field configuration that resembles an hourglass, which is aligned with (but not significantly disturbed by) the bipolar outflow.  In contrast, the fainter source IRS2 shows a magnetic field that has been shaped by the outflow.  The main goal of this discussion is to understand why these differences arise in the two components of the same binary source.

\subsection{Dust polarization and dense-gas tracers}
\label{sec:dense_tracers}

\begin{figure}[hbt!]
\begin{center}
\includegraphics[width=0.49\textwidth, clip, trim=0.3cm 2.35cm -0.4cm 0cm]{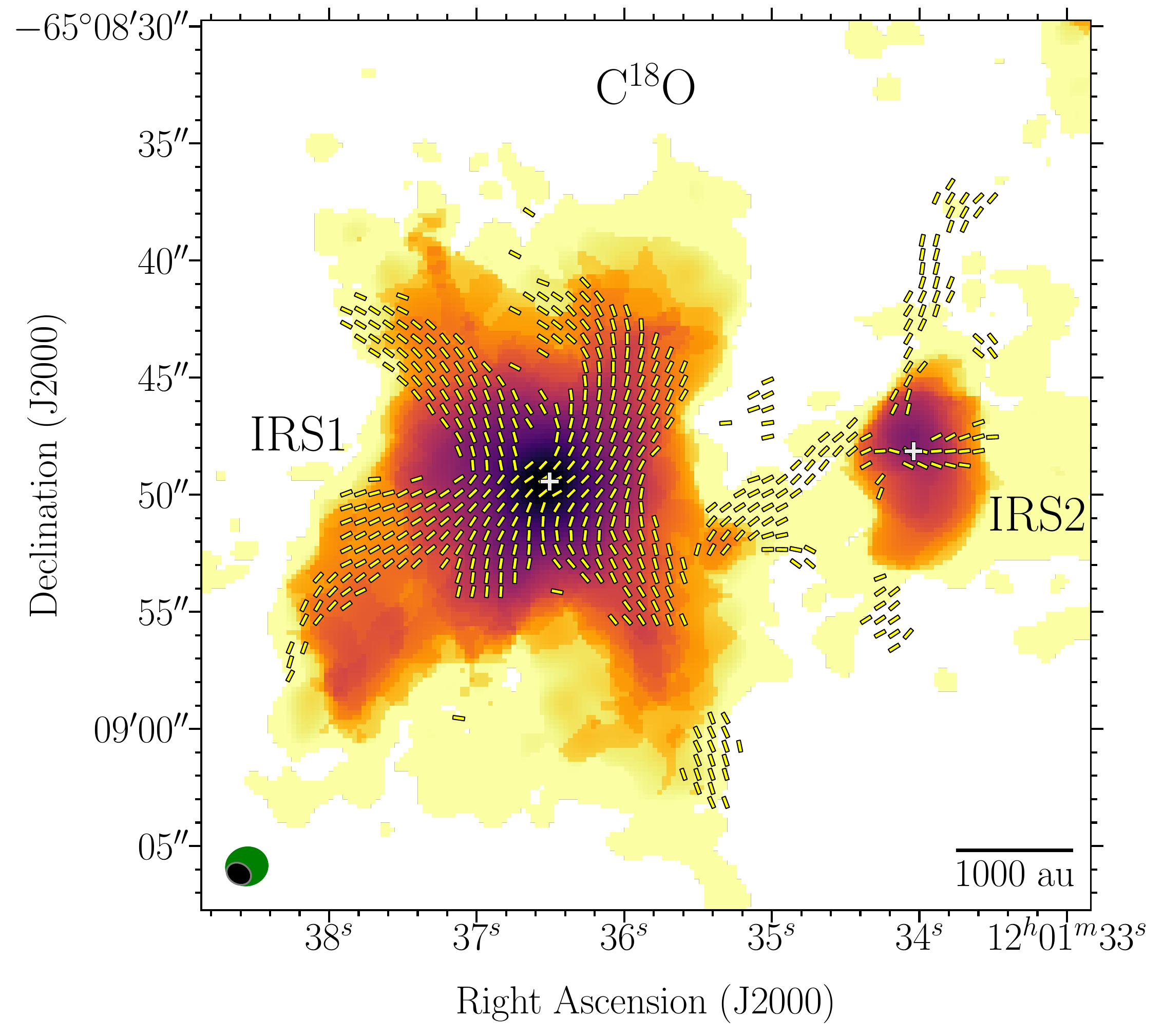}
\includegraphics[width=0.49\textwidth, clip, trim=0.3cm 0cm -0.4cm 0.26cm]{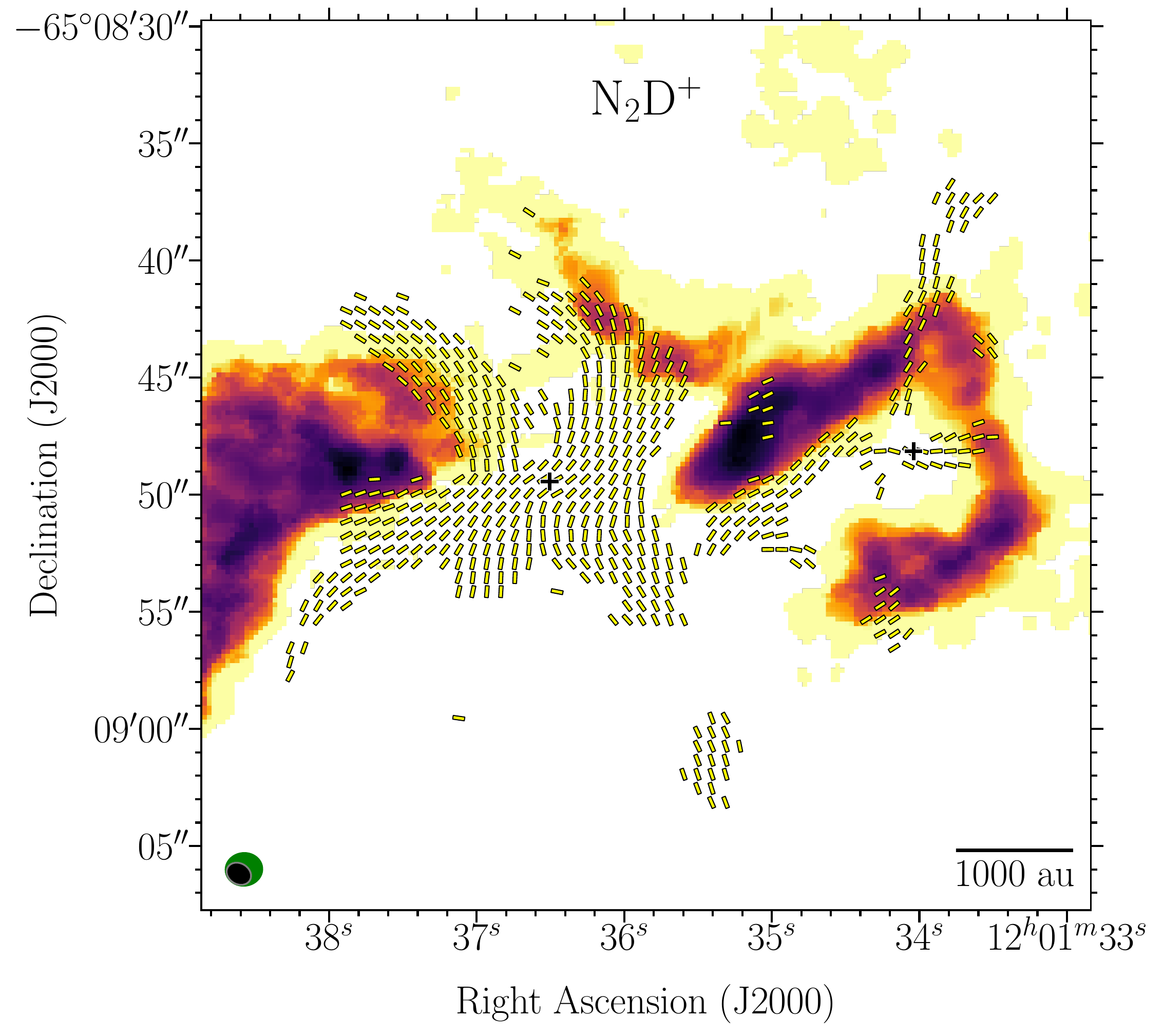}
\end{center}
\vspace{-1em}
\caption{\footnotesize
Maps of integrated (moment 0) \ceighteeno and \ntwodp emission toward \bhr, from the data presented in \tobin.  The C$^{18}$O emission was integrated from --6.7 to --1.7\,\kms; the minimum and maximum values of the map are 0.06 and 3.28\,\jybmkms, respectively.  The N$_2$D$^+$ emission was integrated from --5.5 to --3.4\,\kms; the minimum and maximum values of the map are 0.04 and 0.44\,\jybmkms, respectively.  \bhr has a systemic velocity of --4.5\,\kms \citep{Bourke1997}.  The line segments are the inferred magnetic field orientation, plotted as in Figure~\ref{fig:pol_pfrac}.  The black ellipses in the lower-left corners of both panels represent the synthesized beam of the polarized dust emission, which measures 1$\farcs$10\,$\times$\,0$\farcs$89.  The green ellipses represent the beams of the spectral-line emission; the beam of the C$^{18}$O emission measures 1$\farcs$71\,$\times$\,1$\farcs$54, and the beam of the N$_2$D$^+$ emission measures 1$\farcs$49\,$\times$\,1$\farcs$32.  Crosses indicate the continuum peaks of IRS1 and IRS2.
}
\label{fig:c18o_n2d+}
\end{figure}

\begin{figure*}[hbt!]
\begin{center}
\vspace{1em}
\includegraphics[width=0.54\textwidth, clip, trim=0.3cm 0cm 1.0cm 0cm]{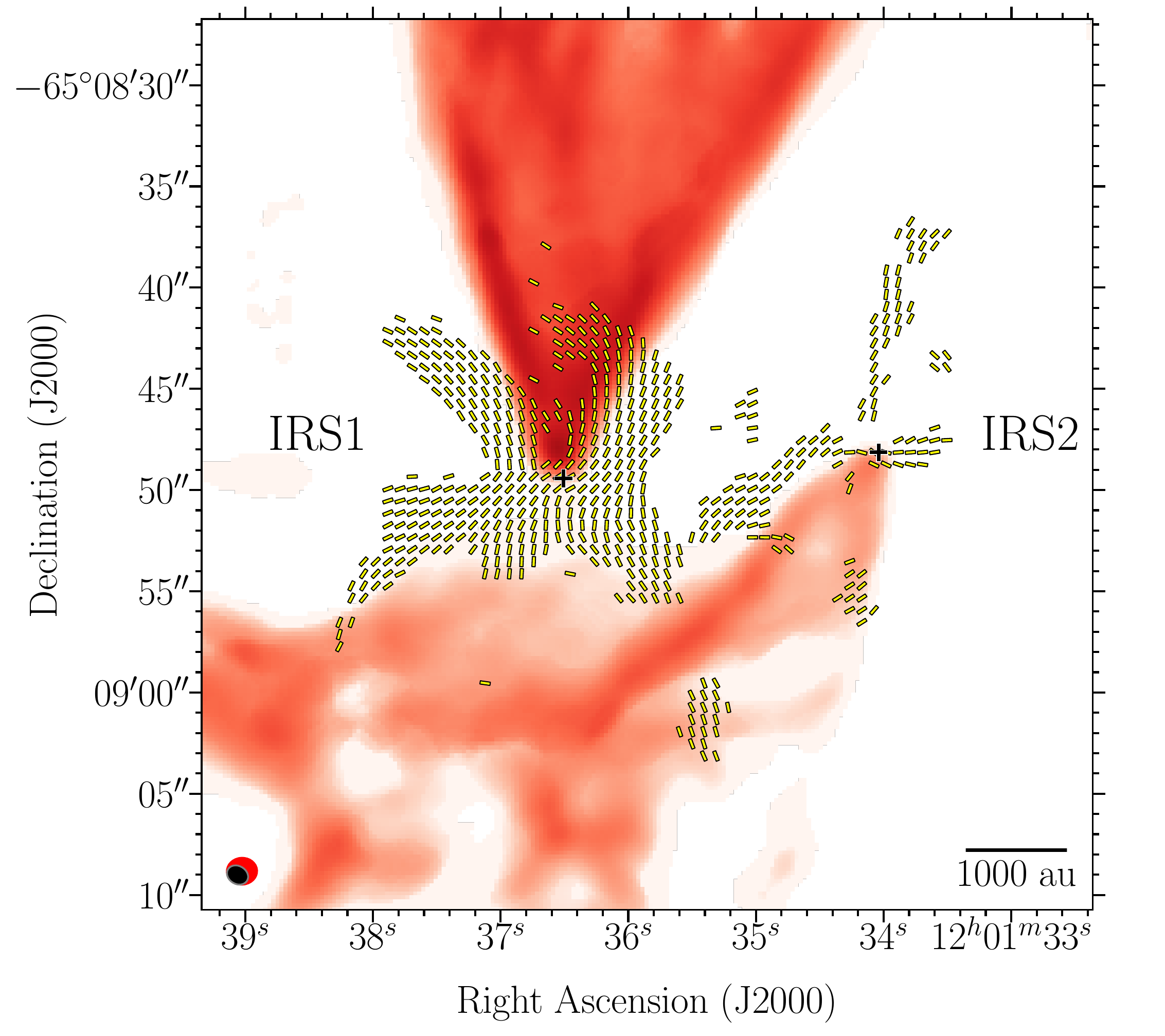}
\includegraphics[width=0.455\textwidth, clip, trim=3.72cm 0cm 1.0cm 0cm]{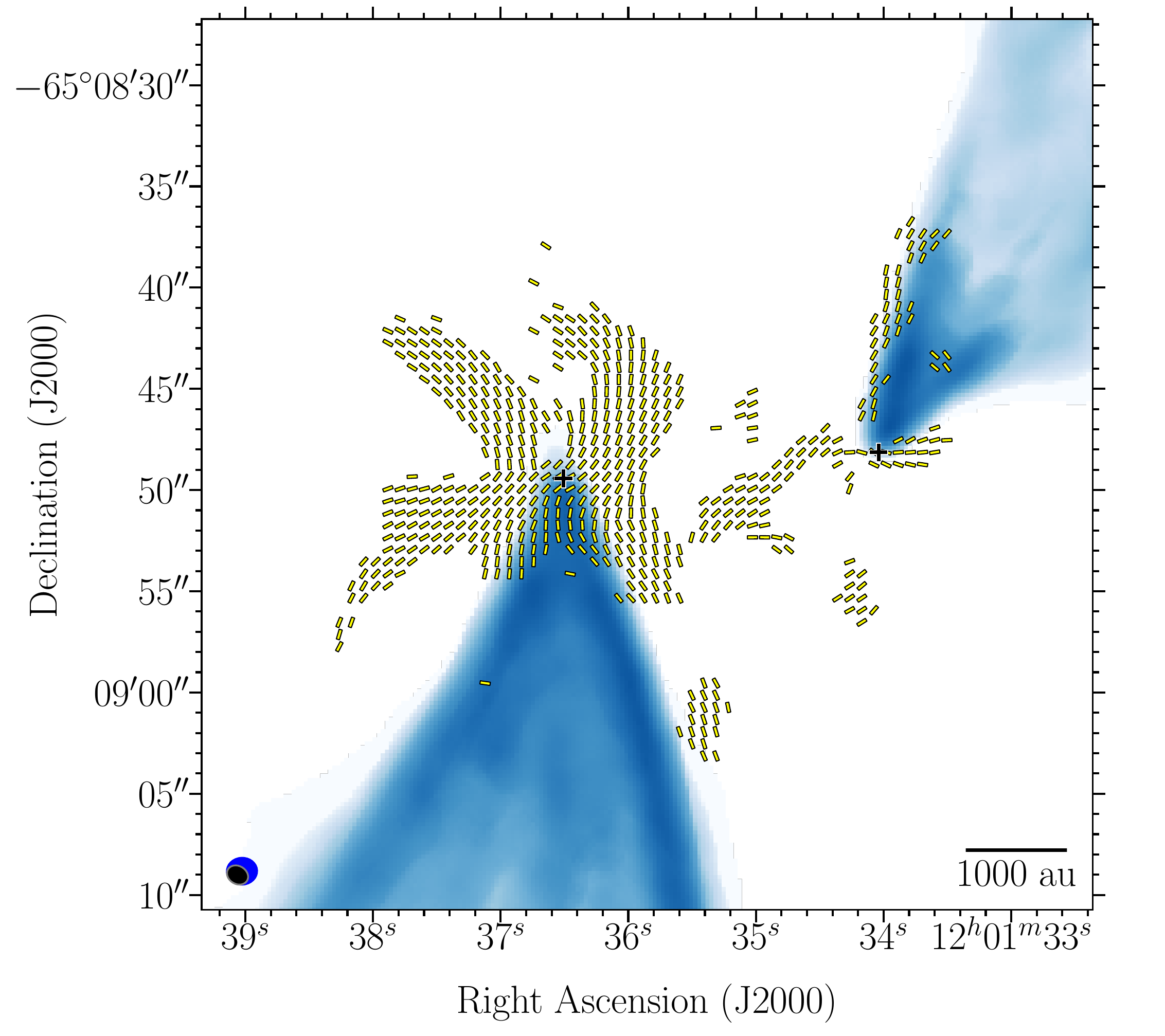}
\end{center}
\vspace{-1em}
\caption{\footnotesize
Maps of integrated (moment 0) \co emission in the red- (left) and blueshifted (right) lobes of the bipolar outflows toward \bhr IRS1 and IRS2, from the data presented in \tobin.  The redshifted emission was integrated from --2.2 to 29.9\,\kms; the minimum and maximum values of the map are 0.6 and 13.05\,\jybmkms, respectively.  The blueshifted emission was integrated from --40 to --6.0\,\kms; the minimum and maximum values of the map are 0.6 and 13.92\,\jybmkms, respectively.  \bhr has a systemic velocity of --4.5\,\kms \citep{Bourke1997}.  The line segments are the inferred magnetic field orientation, plotted as in Figure~\ref{fig:pol_pfrac}.  The black ellipses in the lower-left corners of both panels represent the synthesized beam of the polarized dust emission, which measures 1$\farcs$10\,$\times$\,0$\farcs$89.  The red and blue ellipses represent the beams of the spectral-line emission; the beams of both the blue- and redshifted emission measure 1$\farcs$50\,$\times$\,1$\farcs$33.  Crosses indicate the continuum peaks of IRS1 and IRS2.
\vspace{1em}
}
\label{fig:co}
\end{figure*}

In Figure~\ref{fig:c18o_n2d+} we show the inferred magnetic field toward \bhr overlaid on moment 0 maps of the dense-gas tracers \ceighteeno and \ntwodp emission from \tobin.  The C$^{18}$O and N$_2$D$^+$ are roughly anticorrelated, since the C$^{18}$O traces warm gas in the core and the N$_2$D$^+$ traces cold, pre-stellar gas (see below).  Furthermore, it is clear that the C$^{18}$O emission correlates well with the polarized emission, whereas the N$_2$D$^+$ emission is anticorrelated with the polarization.  These trends are particularly strong in IRS1, which is brighter and warmer than IRS2.    


It is well known that N$_2$D$^+$, and its non-deuterated isotopologue N$_2$H$^+$, are excellent tracers of very cold, pre-stellar material where CO has been frozen out of the gas phase onto dust grains.  CO tends to destroy both N$_2$D$^+$ and N$_2$H$^+$ once the CO sublimates off of the dust grains at a temperature of $\sim$\,25\,K \citep{Aikawa2001, Vasyunina2012, Tobin2013a}.  Furthermore, above the same temperature, the formation pathway of N$_2$D$^+$ is shut off because its precursor molecule H$_2$D$^+$ is no longer being formed, but is rather being converted back into H$_3^+$ and HD \citep{Herbst1982}.  This explains the overall anticorrelation between C$^{18}$O and N$_2$D$^+$ in \bhr.  It also explains the fact that while significant N$_2$D$^+$ emission is still seen on the outskirts of cooler and less luminous IRS2, warmer and brighter IRS1 shows almost no N$_2$D$^+$ emission toward the central core.  This type of drop in emission toward the centers of the protostellar envelope where CO is abundant has also been seen in N$_2$H$^+$ in \bhr \citep{Chen2008} and other Class 0 sources such as, e.g., L483 \citep{Jorgensen2004b}, L1157 \citep{Kwon2015, Anderl2016}, NGC 1333-IRAS 4A and 4B, and L1448C \citep{Anderl2016}.  

This anticorrelation may prove critical for theoretical and observational studies of dust-grain alignment.  For example, a high-resolution survey of either C$^{18}$O or N$_2$D$^+$ (or N$_2$H$^+$) could be used to predict where dust is most likely to be polarized.  The cold environment traced by N$_2$D$^+$, far from the IRS1 and IRS2 protostellar radiation sources, is likely to be the perfect environment for a dramatic decrease in grain-alignment efficiency.  This sort of behavior has been seen in the center of starless cores \citep[e.g.,][]{Alves2014, Andersson2015}, where the decrease in polarization at the center of the source is most likely due to the lack of an anisotropic radiation field (from either external interstellar UV photons or the internal ``lightbulb'' of the protostar itself), which is required for RATs to align dust grains with respect to the magnetic field.

C$^{18}$O is a dense-gas tracer sensitive to high column densities, but only in warm ($\gtrsim$\,25\,K) regions where CO is in the gas phase.  The fact that the polarization closely follows the \ceighteeno emission, particularly toward IRS1, may be because both polarization and C$^{18}$O can be associated with warm regions: CO because it is no longer frozen onto dust grains and has been sublimated into the gas phase; and polarization because the dust is warm, and thus bright, allowing the detection of polarization at the few-percent level.  

Note that while \ceighteeno is spatially coincident with nearly all of the polarization toward IRS1, we only see strong \ceighteeno emission toward the very center of IRS2, and not extended along the outflow lobes.  This suggests that the polarization in IRS1 originates from warm material throughout the natal clump, allowing us to detect its well ordered, hourglass-shaped magnetic field.  However, in IRS2, the polarization originates almost exclusively in the outflow cavity walls, which are strongly irradiated (thus yielding strong polarization), but which lie far away from the central heating source of IRS2 (resulting in a lack of C$^{18}$O, which is frozen out of the gas phase onto the dust grains at large distances from the protostar).

Higher temperatures and stronger irradiation could explain why polarization is so widespread and easily detected in high-mass star-forming regions \citep[e.g.,][]{Zhang2014} and in bright low-mass sources (e.g., \bhr IRS1, NGC 1333-IRAS 4A, IRAS 16293, and L1157), which are warmer and have stronger radiation fields than their lower-luminosity counterparts like IRS2.  In Sections \ref{sec:hourglass_vs_outflow} and \ref{sec:irradiation}, we  discuss the differences in magnetic field between \bhr IRS1 and IRS2 in this context, focusing in particular on the question of irradiation.

\subsection{A natal hourglass in IRS1 versus an outflow-shaped magnetic field in IRS2}
\label{sec:hourglass_vs_outflow}

Despite the powerful outflow emanating from IRS1, the polarization toward IRS1 is not obviously shaped by the outflow.  Rather, the majority of the polarized emission toward the source lies outside of the region of influence of the outflow, which can be see in \co emission in Figure~\ref{fig:co}.  On the eastern edge of IRS1, the two lanes of polarization extending NE and SE of the central source have an orientation that is 30--45$\degree$ different from that of the outflow edge.  This situation is similar to observations of hourglass-shaped magnetic fields in IRAS 4A, IRAS 16293A, and L1157, mentioned in Section \ref{sec:intro}: all of them have powerful bipolar outflows (or two outflows, in some cases), and yet none of them show significant shaping of the magnetic field by the outflows. 

This situation may arise simply because of the high luminosity of these iconic sources.  All of them almost certainly have irradiated outflow cavity walls; however, the clear morphology of the magnetic field along the cavity walls may be obscured because polarized emission is emanating not just from the cavity walls, but from throughout the majority of the envelope, thanks to the strong temperature gradients and to the high fluxes that enable the alignment of dust grains with respect to the magnetic field.  This abundance of aligned grains enables us to detect the natal hourglass-shaped field, which has been preserved from the sources' earlier formation stages.

\begin{figure}[hbt!]
\begin{center}
\includegraphics[width=0.5\textwidth, clip, trim=0.3cm 0cm 0cm 0cm]{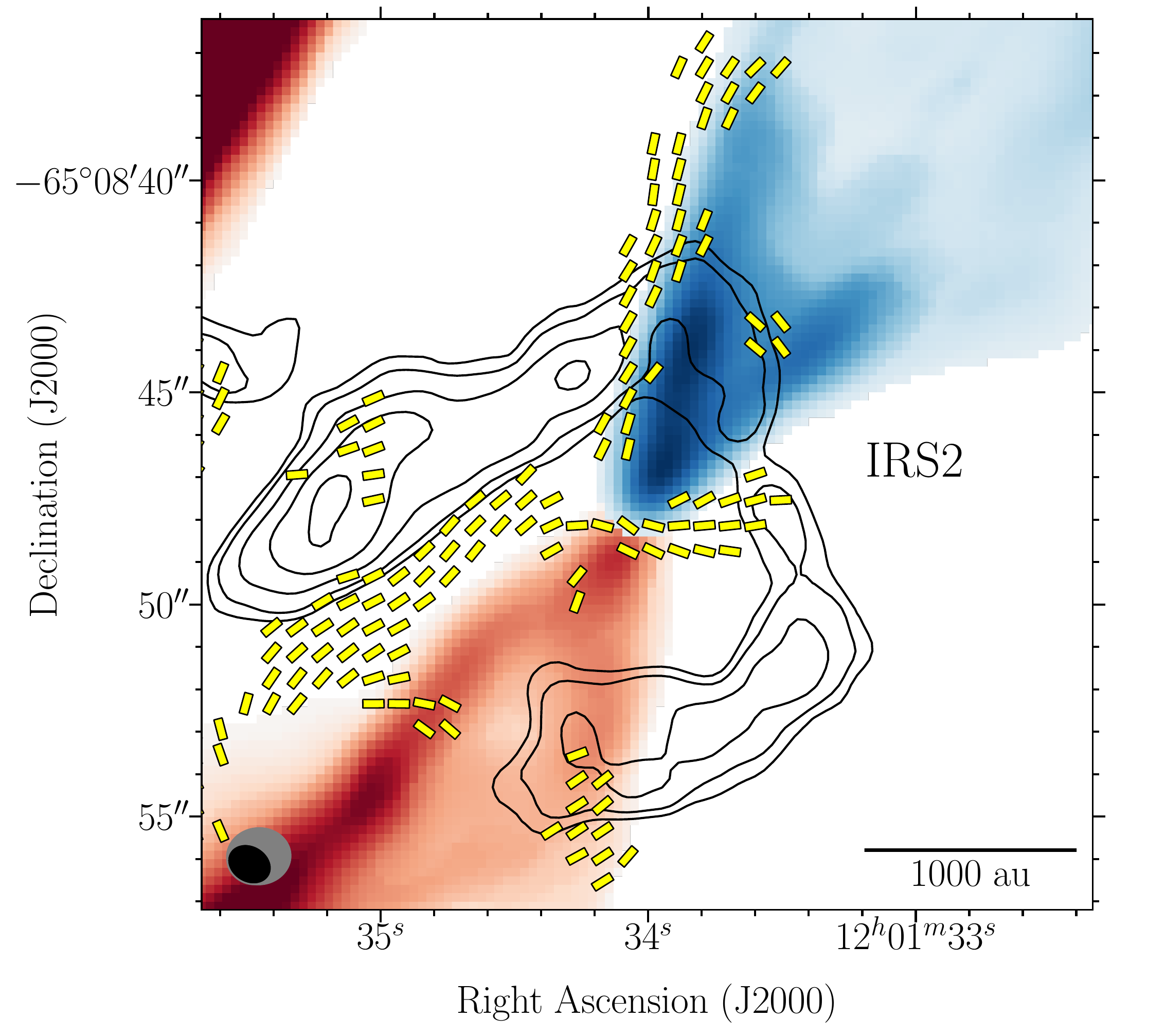}
\end{center}
\vspace{-1em}
\caption{\footnotesize
Zoom-in on IRS2, showing the same \co emission and inferred magnetic field orientation as in Figure~\ref{fig:co}. (Note that the two lobes of the outflow have been plotted on different flux scales to enhance their visibility; as a result, relative fluxes should not be inferred from this image.)  The overplotted contours show the \ntwodp emission at levels of 3,\,4,\,6,\,8,\,10 $\times$ 0.04\,\jybmkms, which is the rms noise level in the moment 0 map (see Figure~\ref{fig:c18o_n2d+}, bottom panel).  The black ellipse represents the synthesized beam of the polarized dust emission, which measures 1$\farcs$10\,$\times$\,0$\farcs$89.  The gray ellipse represents the beams of both the blue- and redshifted CO emission (and is nearly identical to the beam of the N$_2$D$^+$ emission), and measures 1$\farcs$50\,$\times$\,1$\farcs$33.  It is clear that on the redshifted side of the IRS2 outflow, the polarized emission is sandwiched between the outflow emission and the cold, unpolarized material traced by N$_2$D$^+$.
\vspace{1em}
}
\label{fig:irs2_zoom}
\end{figure}

In contrast, less luminous sources like \bhr IRS2, B335 \citep{Maury2018}, and Ser-emb 8(N) \citep{LeGouellec2019a} all show polarization that clearly has been affected by the outflow.  When we examine the polarization map of IRS2 in detail, we also see that the polarization along the northern edge of the redshifted outflow lobe lies between the outflow and a cold, dense streamer of N$_2$D$^+$ (see Figure~\ref{fig:irs2_zoom}).  The fact that the polarization is sandwiched between the outflowing gas and the unpolarized material traced by N$_2$D$^+$ suggests that the origin of the enhanced polarization is the irradiation of the outflow cavity walls---but only to a depth of $\sim$\,300\,au, beyond which the dust is cold and unpolarized.  We suspect that this is the case for all of these fainter sources, which have neither the strong temperature gradients nor the high fluxes of the brighter, aforementioned sources.  Consequently, toward these faint sources we are only able to detect polarization in regions with enhanced polarization, such as in the outflow cavity walls.

\subsection{Irradiation of the outflow cavity walls in \bhr IRS2}
\label{sec:irradiation}

Single-dish studies of high-$J$ CO emission toward protostars have suggested that such energetic CO emission cannot be reproduced by models of passive envelope heating.  Rather, this gas must be heated by UV radiation originating in the accretion shock around the central protostar as well as in shocks distributed throughout the outflow itself \citep{Spaans1995, vanKempen2009a, Visser2012}.  More recent models explained the shape of the high-$J$ CO ladder and the chemical signatures (e.g., the line ratios H$_2$O/CO and H$_2$O/OH) seen in \textit{Herschel} observations of embedded protostars by invoking models of UV-irradiated shocks in the protostars' bipolar outflow cavities \citep{Kristensen2017, Karska2018}.  These findings are consistent with many other observational and theoretical studies that have seen enhanced chemistry in the vicinity of shocked, irradiated gas \citep[e.g., in HH objects:][]{Girart1994, Taylor1996, Viti1999, Girart2002, Christie2011}.  \cosixtofive was detected in observations of \bhr with the Atacama Pathfinder EXperiment (APEX) telescope by \citet{vanKempen2009c} and \citet{Gusdorf2015}; the observations by \citeauthor{Gusdorf2015} resolve this high-$J$ CO emission in both the IRS1 and IRS2 outflows.  Here we consider the scenario that the enhanced polarization along the outflow cavity walls of IRS2 may be the result of irradiation by photons generated in accretion and outflow shocks.

The polarized lane along the northern edge of the blue- and redshifted IRS2 outflow lobes is marginally resolved, with an average thickness of $\sim$\,300\,au.  Based on PDR models by \citealt{Girart2005}, UV radiation is fully extincted at an $A_V$ of order unity, which is achieved at a molecular hydrogen column density of between 10$^{21}$ and 10$^{22}$\,cm$^{-2}$ \citep{Bohlin1978}.  
To estimate the amount of material along the IRS2 outflow cavity wall, we use the standard conversion from flux density $S_{\nu}$ to gas mass $M_{\mathrm{gas}}$ \citep[see, e.g.,][]{Hull2017b, HullZhang2019}:

\begin{equation}
M_{\mathrm{gas}}=\frac{S_{\nu}d^{2}}{\kappa_{\nu}B_{\nu}\left(T_{\mathrm{d}}\right)}\,\,,
\label{eq:M_gas}
\end{equation}

\noindent
where $B_{\nu}\left(T_{\mathrm{d}}\right)$ is the Planck function at the 233\,GHz frequency of our observations, the distance $d=200$\,pc, and the dust opacity at 1.3\,mm $\kappa_{\nu} = 2\,\mathrm{cm}^2/\mathrm{g}$ \citep{Ossenkopf1994}.  We assume a gas-to-dust ratio of 100.     

In a circle with a diameter equal to 300\,au, which is the approximate width of the enhanced dust emission along the outflow cavity, the flux density along the entire northern edges of the blue- and redshifted outflow cavity walls is a roughly constant $\sim$\,2--3\,mJy.  The fact that the brightness of the cavity walls is constant---as opposed to decreasing with distance from the central protostar---suggests that the heating of the outflow cavities is indeed from distributed shocks and not purely from high-energy radiation from the central protostar, the latter of which would decrease with distance from the central source.  This is perhaps unsurprising in the case of IRS2, which, in addition to having a bipolar CO outflow, is known to have a high-velocity ($\pm$\,100\,\kms), bipolar SiO jet, whose internal shocks may contribute to the illumination of the cavity walls (Bourke et al., in prep.).

To calculate the column density along the IRS2 outflow cavity walls, we use the IRS2 temperature estimate of 20\,K from \tobin, assume a mean molecular weight of 2.8 in the gas \citep{Kauffmann2008}, and assume an average flux of 2.5\,mJy in a 300\,au-diameter circle centered on the outflow cavity wall.  The resulting column density is $\sim$\,$2.6 \times 10^{22}$\,cm$^{-2}$.  However, the dust along the outflow walls could be warmer than the overall dust temperature toward the IRS2 core.  The \textit{gas} temperature of the IRS2 outflow could be up to 300\,K \citep{Parise2006}, which is reasonable based on the models of UV-irradiated outflow cavities by, e.g., \citet{Visser2012} and \citet{Drozdovskaya2015}.  If we take that same value as the \textit{dust} temperature (which is perhaps unreasonably high, as the dust is usually significantly cooler than the gas: \citealt{Yildiz2015}), the column density we derive is $\sim$\,$1.3 \times 10^{21}$\,cm$^{-2}$.  Even considering this high value of 300\,K for the dust temperature, the column density of the cavity walls is still too high for UV photons to penetrate to a depth of 300\,au.  In light of this, below we paint two pictures of the potential origin of the polarization we see in the outflow cavity walls in IRS2: the ``thick'' scenario, and the ``thin'' scenario.

We first consider the ``thick'' scenario, where the polarization originates in a layer of the cavity walls with a thickness of $\sim$\,300\,au.  The first question to be addressed is, Which photons can penetrate to that depth and align the dust grains?  We thus calculate how far into the walls photons with wavelengths longer than those of UV photons can penetrate.  We use the dust opacity values $\kappa_\nu$ from \citet{Ossenkopf1994} that correspond to gas densities of $10^6$\,cm$^{-3}$ (and thus dust densities of $10^4$\,cm$^{-3}$, assuming a gas-to-dust ratio of 100) and grains without icy mantles.\footnote{The gas number density in the cavity walls of IRS2 is on the order of $10^{7}$\,cm$^{-3}$, assuming a dust temperature of 20\,K.  We calculated this value by dividing the mass from Equation \ref{eq:M_gas} by the volume of a sphere with a diameter of 300\,au, the same as the approximate thickness of the cavity walls.  This density is between the gas density values of $10^6$\,cm$^{-3}$ and $10^8$\,cm$^{-3}$ for which $\kappa_\nu$ is calculated by \citet{Ossenkopf1994}; however, the values of $\kappa_\nu$ for the two densities differ by less than a factor of two at all relevant frequencies.}
Assuming a constant opacity $\kappa_\nu$ and mass density $\rho$ of the dust grains throughout the thickness of the cavity wall, the path length $s$ to the point where the optical depth $\tau = 1$ can be calculated simply as $s = 1 / \kappa_\nu \rho$.  The values of $\kappa_\nu$ in \citet{Ossenkopf1994} are normalized by the dust mass density $\rho$, hence we must multiply $\kappa_\nu$ by the value of $\rho$ in the cavity walls in order to calculate $s$.

For 1\,$\micron$ photons, the penetration depths assuming 20\,K and 300\,K temperatures are $\sim$\,14\,au and $\sim$\,280\,au, respectively.
For 10\,$\micron$ photons, the depths are $\sim$\,70\,au and $\sim$\,1400\,au.
For 100\,$\micron$ photons, the depths are $\sim$\,2800\,au and $\sim$\,56,000\,au.
Finally, for 1.3\,mm photons, the depths are $\sim$\,84,000\,au and $\sim$\,1,700,000\,au.
If indeed the polarization originates in a layer with a thickness of $\sim$\,300\,au, then it appears that the dust grains have been aligned by mid- to far-infrared (MIR/FIR) photons and/or possibly by an anisotropic radiation field resulting from the temperature gradient between the inner part of the outflow cavity and the cold, unpolarized material traced by \ntwodp.

A major caveat of this ``thick'' scenario is grain growth.  RAT theory suggests that photons can efficiently spin-up dust grains only if the grain size is comparable to the photon wavelength \citep{LazarianHoang2007a}.  Therefore, in order for the MIR/FIR photons to be the cause of the enhanced polarization we see in IRS2, there would need to be a substantial population of $\sim$\,10\,$\micron$-sized dust grains in the cavity walls.  While grains of that size are expected in circumstellar disks based on both SED-fitting and dust-scattering-polarization studies \citep[e.g.,][]{Perez2012, Testi2014, Stephens2017b, Hull2018a}, grain growth in the regions where we see polarization toward IRS2 here---at distances that in some places exceed $\sim$\,1000\,au from the central source---is somewhat surprising.  Note, however, that there have been hints of grain growth in Class 0 and Class I protostellar cores from millimeter-wavelength observations \citep{Jorgensen2007, Kwon2009, Chiang2012, Miotello2014, LiJ2017, Galametz2019}.

\begin{figure*}[hbt!]
\vspace{0em}
\begin{center}
\includegraphics[width=0.361\textwidth, clip, trim=0cm 0cm 1cm 0cm]{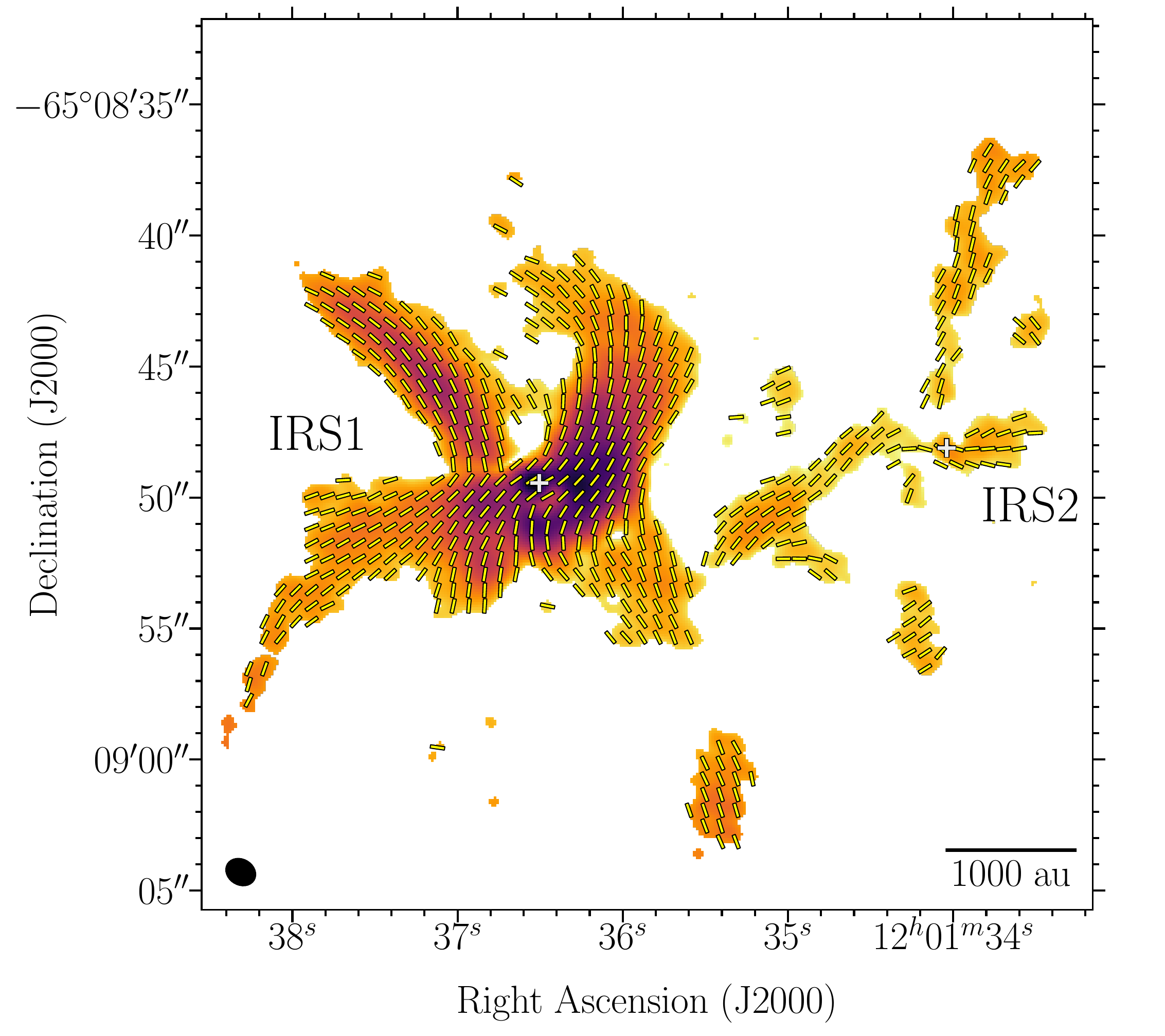}
\includegraphics[width=0.30\textwidth, clip, trim=3.73cm 0cm 1cm 0cm]{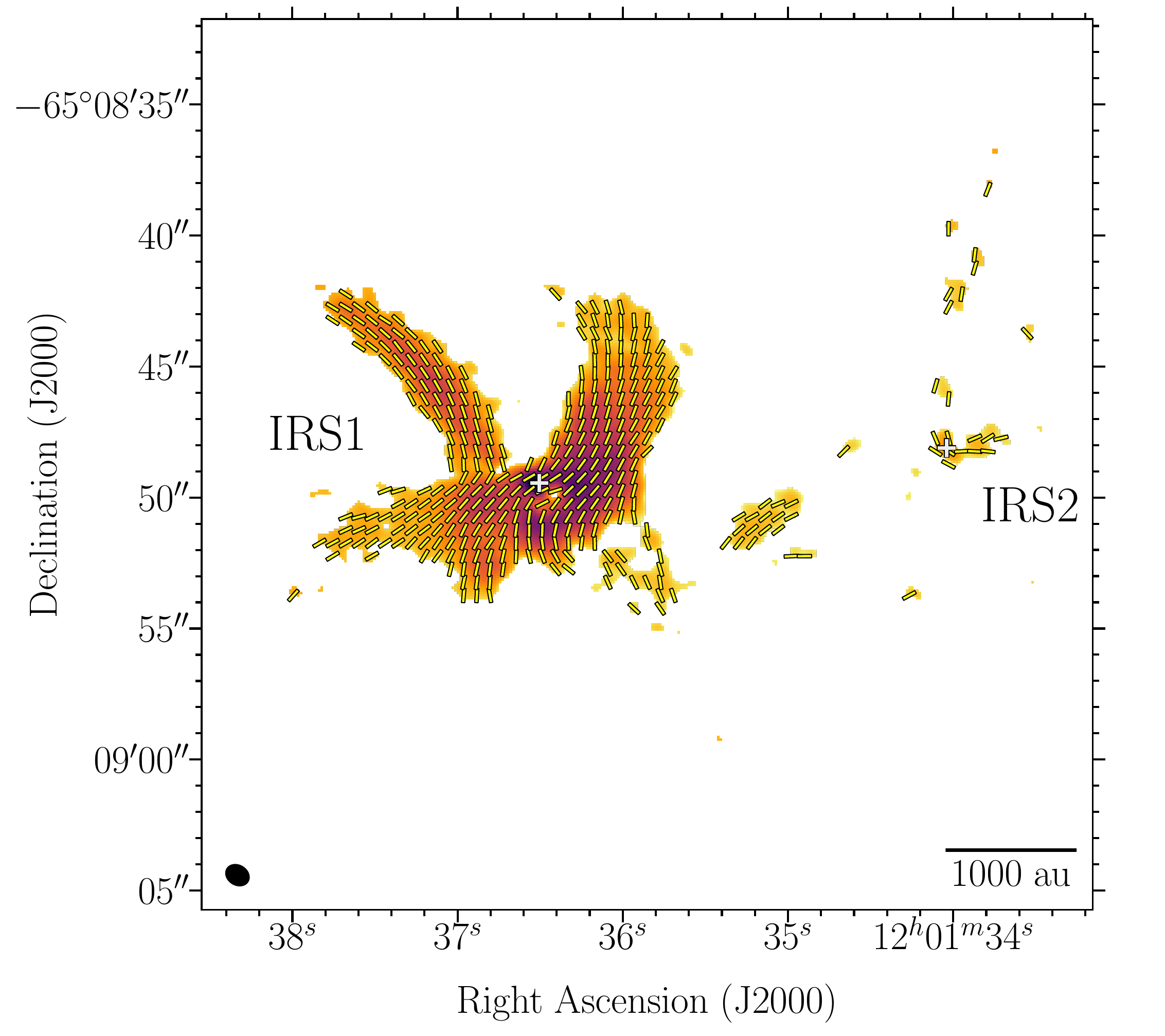}
\includegraphics[width=0.30\textwidth, clip, trim=3.73cm 0cm 1cm 0cm]{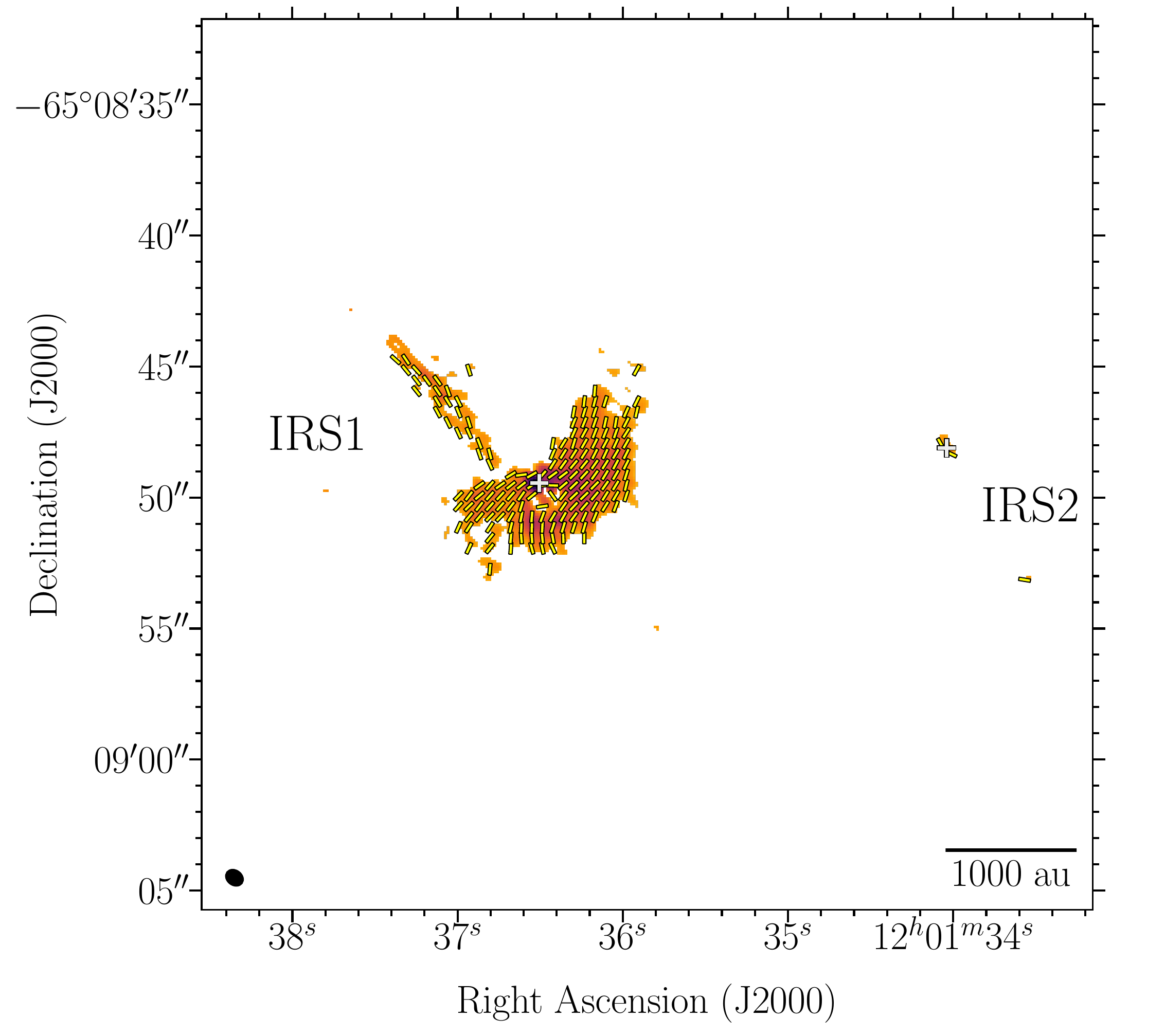}
\end{center}
\vspace{-1em}
\caption{\footnotesize
Multi-resolution images of the polarized intensity $P$ (color scale) made using \texttt{robust}\,=\,2.0 (left), 0.5 (middle), and --0.5 (right), where the respective synthesized beam sizes are 1$\farcs$10\,$\times$\,0$\farcs$89, 0$\farcs$85\,$\times$\,0$\farcs$67, and 0$\farcs$63\,$\times$\,0$\farcs$48.  Line segments are the inferred magnetic field orientation toward \bhr IRS1 and IRS2, plotted as in Figure~\ref{fig:pol_pfrac}, where $\sigma_P$\,=\,25\,\ujybm for \texttt{robust}\,=\,2.0 and 0.5, and 40\,\ujybm for \texttt{robust}\,=\,--0.5.  Crosses indicate the continuum peaks of IRS1 and IRS2. The sharp, magnetized feature to the NE of IRS1 is enhanced with increasing resolution.
\vspace{1em}
}
\label{fig:3res}
\end{figure*}

We also consider a second, ``thin'' scenario, where the resolved layer of polarization that we see is caused by projection effects.  In this case, the observed polarization actually originates from a very thin layer of material in the cavity wall, which comprises a negligible amount of mass, but which is distributed around a substantial fraction of the diameter of the cavity where the column density of dust is still high enough to detect polarized emission.  This scenario would allow higher-energy photons (i.e., UV, optical, NIR), which cannot penetrate very deep into the cavity walls (see above), to align the smaller, micron-sized dust grains that are expected to dominate the grain population at large distances from the central source.  This thin layer of aligned, small grains could then be co-located with molecular emission associated with UV-driven chemistry in outflow cavities \citep[][and references therein]{Karska2018}; emission from several UV tracers has been spatially resolved in ALMA observations \citep[e.g.,][]{Imai2016, LeGouellec2019a}, but is expected to originate mainly in the very thin layer of the cavity walls accessible to the UV radiation.

This ``thin'' scenario has its own caveats.  If only an extremely thin layer of dust grains were aligned, it seems unlikely that the polarization fraction would reach the high (>\,10\%) levels that we detect in IRS2, considering that the polarization fraction is derived from the polarized intensity from the aligned grains divided by the total intensity of the dust emission, the latter of which originates from the entire core.  In addition, the high-energy radiation would be reprocessed by the thin layer of initially-heated dust grains in the cavity wall, being re-emitted as longer-wavelength photons that could penetrate deeper into the walls.  However, it is unclear if there would be enough of these photons to align a significant number of grains to an appreciable depth beyond the initial heated layer.

Given the major caveat of the polarization fraction levels in the ``thin'' case, and the fact that recent studies have shown that grain growth in Class 0 protostellar cores may be necessary to explain high-resolution, interferometric polarization results \citep{Maury2018, Valdivia2019, LeGouellec2019a}, we find the ``thick'' scenario to be the more plausible of the two possibilities for the origin of the polarization in the IRS2 outflow cavities.  However, the answer is far from clear.  Future, multi-wavelength polarization observations of protostellar cores, coupled with synthetic observations of MHD simulations with next-generation radiative transfer codes like POLARIS \citep{Reissl2016} that incorporate dust-grain alignment via RATs, will help us to better understand the origin of polarization in irradiated environments like cavity walls.

\subsection{A possible magnetized accretion streamer in \bhr IRS1}
\label{sec:pol_origins}

The most intriguing emission feature in the $P$ map of \bhr IRS1 is the strip oriented NE of the central source (Figure~\ref{fig:QU_POLI}, bottom panel).  This feature is most clearly seen in the multi-resolution image shown in Figure~\ref{fig:3res}, and is oriented at a very different angle from the eastern edge of the redshifted outflow cavity wall (Figure~\ref{fig:co}).  If it is indeed part of the natal hourglass structure in IRS1, then it is not clear why it is so much sharper and brighter than the other ``corners'' of the hourglass, which are resolved out (or are too faint to be detected) at higher resolution.  This is a similar situation to Serpens~SMM1, where the polarization along the E--W cavity wall seen in \citet{Hull2017b} is not detectable at higher resolution, whereas extremely sharp, bright polarization structures near the N--S wall of the outflow cavity remain visible at higher resolution \citep{LeGouellec2019a}.  These sharp, approximately linear structures south of SMM1-a and the sharp structure to the NE of \bhr IRS1 have similar widths of $\sim$\,50\,au.  

Narrow structures have also been seen in other sources, including filamentary features surrounding the low-mass protostellar core Ser-emb 8 \citep{Hull2017a}; the N--S equatorial-plane feature in B335 \citep{Maury2018}; the ``bridge'' and streamers observed toward the two binary components in IRAS 16293 \citep{Sadavoy2018c}; and the ``arm-like'' structure in OMC3-MMS6 \citep{Takahashi2019}.  The existence of these features, which are always more prominent in $P$ maps than in $I$ maps, suggest that the production of polarized emission inside a protostellar core is strongly dependent on the local environmental conditions such as the optical depth and the anisotropy of the local radiation field.  As discussed in \citet{LeGouellec2019a}, the high polarization of these regions is surprising, considering that they are deeply embedded and far away from any obvious source of strong irradiation.

\begin{figure*}[hbt!]
\begin{center}
\includegraphics[width=1.0\textwidth, clip, trim=4cm 0cm 4cm 0cm]{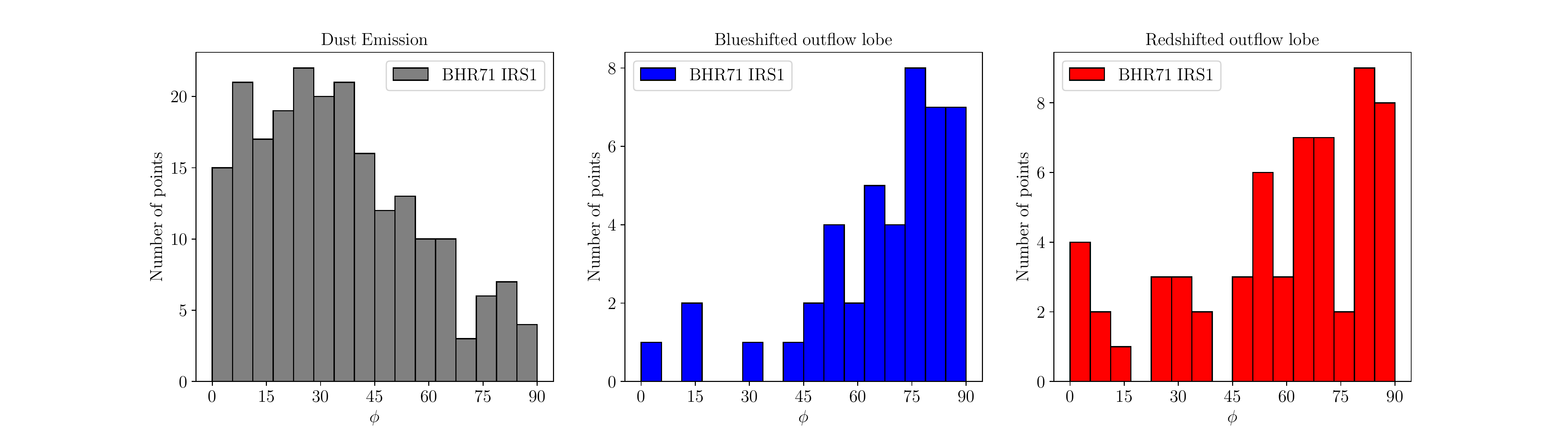}
\includegraphics[width=1.0\textwidth, clip, trim=4cm 0cm 4cm 0cm]{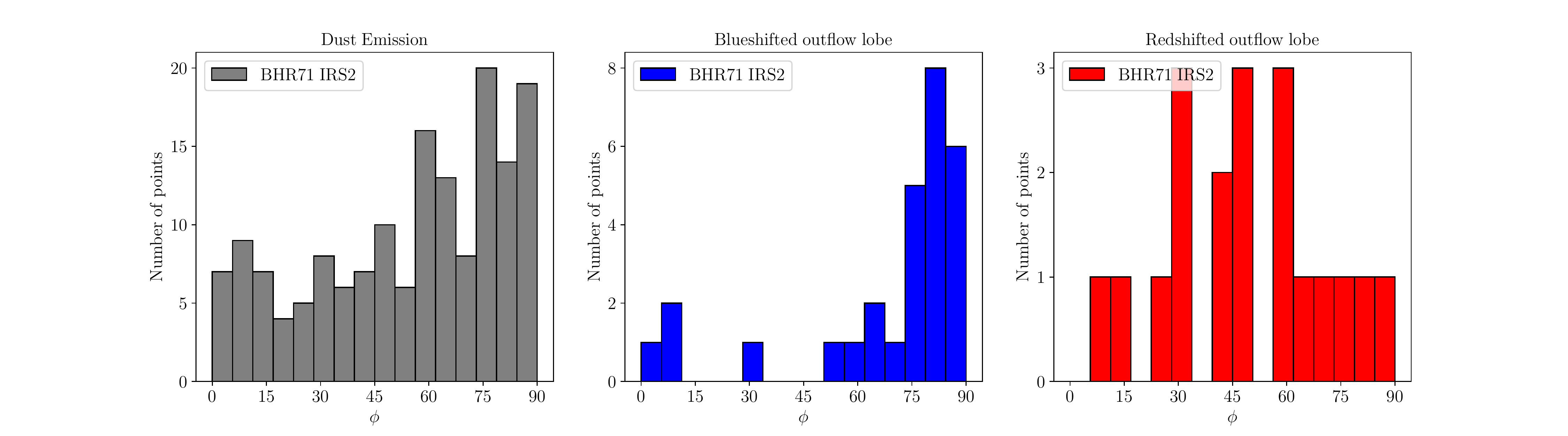}
\end{center}
\vspace{-1em}
\caption{\footnotesize
Histograms of relative orientation (HROs) of $\phi$, defined as the difference between the magnetic field orientation and the Stokes $I$ dust-intensity gradient (left); and the difference between the magnetic field orientation and the gradients in the blueshifted (center) and redshifted (right) outflow emission toward both IRS1 and IRS2.  We calculate the gradient in the zone of significant signal-to-noise: in the Stokes $I$ continuum maps where the signal $I > 3$\,$\sigma_I$, and in moment 0 maps produced with a threshold at 10\,$\times$ the rms noise level in a channel of the CO image cube (the moment 0 maps we analyze are the same as those plotted in Figure~\ref{fig:co}).  We then select high gradient values (see \citealt{LeGouellec2019a} for details). The angles between the magnetic field and the gradients are calculated $\sim$\,4\,$\times$ per synthesized beam (Nyquist, or twice, in both RA and DEC), where the beam width is $\sim$\,0\farcs5 for the dust maps and $\sim$\,0\farcs7 for the CO maps.
\vspace{1em}
}
\label{fig:hro}
\end{figure*}

It is conceivable that these sharp, filamentary features are the result of grains aligned in accretion streamers; see \citet{Alves2019} for the highest-resolution observations to date of accretion streamers that are funneling material onto the central sources in a low-mass protostellar binary.  Theoretical work by \citet{LazarianHoang2007b} and \citet{Hoang2018} on grain alignment via mechanical alignment torques (MATs) has suggested that helical grains could be mechanically 
aligned\footnote{Note that this is different from the traditional ``Gold alignment'' version of mechanical alignment \citep{Gold1952}.  Gold alignment produces polarization oriented along the direction of the flow, and thus would yield an inferred magnetic field along the minor axis of the filament, whereas we see a magnetic field aligned with the filament's major axis.} 
with the magnetic field by the drift (both sub- and supersonic) of gas relative to dust grains.  These relative flows of gas and dust could occur in an 
outflow,\footnote{Note that while the polarization toward IRS2 is strongly associated with the outflow cavity walls, we find it unlikely that the action of the outflow itself is producing the polarization via MATs.  The main argument against MATs in this case is that the northern edge of the redshifted outflow lobe from IRS2 is significantly offset from the observed polarized dust emission.  We therefore find it more likely that the enhanced polarization toward IRS2 is due to irradiation of the outflow cavity walls.}
an accretion inflow/streamer, or possibly in a region of strong ambipolar diffusion between the weakly charged dust and the neutral gas.  The MAT mechanism yields a polarization orientation consistent with what we see toward the highly polarized filament to the NE of IRS1.  However, the kinematics of the \ntwodp and \ceighteeno emission from \tobin do not reveal any such accretion features, and thus it is not possible to confirm this scenario with our current suite of dense-gas tracers.

\subsection{Histograms of Relative Orientation of the magnetic field versus dust and outflow emission in \bhr IRS1 and IRS2}

To better understand the differences in the magnetic field morphologies of IRS1 and IRS2, we analyze the Histogram of Relative Orientation (HRO; \citealt{Soler2013}) of the magnetic field versus the gradients of both dust and the \co (outflow) emission toward IRS1 and IRS2 (see Figure~\ref{fig:hro}).  HROs have been used many times recently to shed light on the importance (or lack thereof) of the magnetic field in the formation of structure in star-forming regions, from the spatial scales of molecular clouds \citep{PlanckXXXII, PlanckXXXV, Soler2017, Fissel2019, Soler2019} to individual protostellar cores \citep{Hull2017a}.

When producing the histograms, both sources were isolated to clearly establish the corresponding distributions. We calculated the gradient in the zone of significant signal-to-noise: in the Stokes $I$ continuum map where the signal $I > 3$\,$\sigma_I$, and in moment 0 maps produced with a threshold at 10\,$\times$ the rms noise level in a channel of the CO image cube (the moment 0 maps we analyze are the same as those plotted in Figure~\ref{fig:co}).  We then selected the zones with significant gradient values.  In the outflow moment 0 maps this served to highlight the edges of the outflow cavities, whereas in the dust continuum maps the gradient picked out both the central cores of the protostars as well as (in the case of IRS2) regions of enhanced dust emission along the outflow cavity walls.  Finally, for the locations where there is both a gradient value as well as a magnetic field orientation (i.e., where $P > 3$\,$\sigma_P$), we derived $\phi$, defined as the difference in angles between the magnetic field and the emission gradient, and added the points to the distributions.  For further details of this HRO analysis, see \citet{LeGouellec2019a}.

Toward IRS2, it is clear from Figure~\ref{fig:irs2_zoom} that the magnetic field follows the northern edge of the outflow cavity.  The HRO comparing the magnetic field and the redshifted CO emission does not exhibit this, as it is limited by statistics because of the physical offset between the redshifted outflow lobe and the polarized emission.  However, we can clearly see that in the blueshifted lobe of IRS2, $\phi$ peaks near 90$\degree$, indicating that the magnetic field orientation tends to be perpendicular to the CO emission gradient (i.e., nearly parallel to the edges of the outflow).  For IRS1, however, the HRO for both the blue- and redshifted lobes look similar to the blueshifted lobe of IRS2, simply because the hourglass symmetry axis and the outflow axis are aligned. The curvature of the hourglass-shaped magnetic field in IRS1 does yield a broader HRO compared with the HRO toward the blueshifted outflow lobe of IRS2; however, this is not a strong distinguishing factor.  The fact that the hourglass magnetic field in IRS1 and the magnetic field along the edges of the outflow in IRS2 yield such similar distributions in the magnetic-field-versus-CO HROs highlights the difficulty of distinguishing between the ``natal hourglass'' versus ``outflow-affected magnetic field'' scenarios.  

We can make a clearer distinction between the two sources by looking at the HRO of the magnetic field versus the dust emission (Figure~\ref{fig:hro}, left-hand panels). The HRO from IRS2 demonstrates that the magnetic field is perpendicular to the dust-emission gradient (i.e., the magnetic field is parallel to the outflow-cavity walls), indicating that the magnetic field has been affected by the outflow.  In contrast, the HRO from IRS1 shows that the field is more parallel to the dust intensity gradient, indicative of a magnetically regulated but gravity-dominated scenario in a centrally condensed protostellar core \citep[see, e.g.,][]{Koch2012, Koch2018}.  These differences in the magnetic-field-versus-dust HROs can help us distinguish whether the dust morphology in a source has been more affected by gravity or by the outflow; however, it is still essential to find a robust way to use \textit{outflow} tracers to determine quantitatively whether magnetic fields have been affected by outflows.

\section{Conclusions}
\label{sec:con}

We have presented 1.3\,mm ALMA observations of polarized dust emission toward the wide-binary protostellar system \bhr.  After analyzing the inferred magnetic field morphology toward both sources alongside maps of the bipolar outflows and dense-gas tracers, we come to the following conclusions:

\begin{enumerate}

\item While the magnetic field morphologies of both \bhr IRS1 and IRS2 are consistent with hourglass shapes, analysis of the magnetic field maps alongside spectral-line observations reveals that IRS1 has what appears to be a natal, hourglass-shaped magnetic field.  In contrast, its fainter, more embedded binary counterpart IRS2 exhibits a magnetic field that has been affected by its bipolar outflow.

\item Toward IRS1, there is a strong correlation of polarized emission with C$^{18}$O, which traces warm ($\gtrsim$\,25\,K) material throughout the whole protostellar envelope.  Toward IRS2, in contrast, the polarization is confined mainly to the outflow cavity walls.  

\item Along the northern edge of the redshifted outflow cavity in IRS2, the polarized emission is sandwiched between the outflowing material and a filament of cold, dense gas traced by N$_2$D$^+$, toward which no dust polarization is detected.  This suggests that the origin of the enhanced polarization in IRS2 is the irradiation of the outflow cavity walls, which enables the alignment of dust grains with respect to the magnetic field---but only to a depth of $\sim$\,300\,au, beyond which the dust is cold and unpolarized.  However, in order to align grains deep enough in the cavity walls, and to produce the high polarization fraction seen in IRS2, the aligning photons are likely to be in the mid- to far-infrared range, which suggests a degree of grain growth beyond what is typically expected in very young, Class 0 sources.

\item The anticorrelation of dust polarization from IRS1 and IRS2 and emission from N$_2$D$^+$ suggests that this species (and its non-deuterated counterpart N$_2$H$^+$) is an excellent tracer of unpolarized material because it is very sensitive to regions of cold, dense gas where Radiative Alignment Torques (RATs) cannot efficiently align dust grains with the magnetic field.

\item The difference in magnetic field morphologies toward the two binary components of \bhr may arise simply because of the higher temperature and $\sim$\,10\,$\times$ higher luminosity of IRS1 relative to IRS2.  The higher temperature yields warmer dust, and thus more easily detectible polarization.  The higher luminosity yields a stronger radiation field and larger temperature gradients, which enable the alignment of dust grains with respect to the magnetic field throughout the majority of the envelope of IRS1.  In contrast, toward less luminous IRS2, we are only able to detect polarization in regions with enhanced polarization, such as the irradiated outflow cavity walls.  This same logic could explain why polarization is so widespread and easily detected in high-mass star-forming regions and in bright low-mass sources (like IRS1), which are warmer and have stronger radiation fields than their lower-luminosity counterparts (like IRS2).

\item Recent ALMA observations have revealed narrow polarization features in sources such as \bhr IRS1 and IRS2 (shown here), Ser-emb 8 \citep{Hull2017a}, Serpens SMM1 \citep{Hull2017b, LeGouellec2019a}, B335 \citep{Maury2018}, IRAS 16293 \citep{Sadavoy2018c}, OMC3-MMS6 \citep{Takahashi2019}, and Serpens Emb 8(N) \citep{LeGouellec2019a}.  These features seem to fall into two categories:

\subitem (a) \textit{Outflow-related features.}  These manifest themselves as features that have very high polarization fractions (sometimes $\gtrsim$\,20\%), and that are well aligned with the outflow cavity walls.  Examples are the outflow cavities of \bhr IRS2 (shown here); the E--W wall of the redshifted, low-velocity outflow lobe of Serpens SMM1-a; and the outflow cavities toward both B335 and Ser-emb 8(N).  The polarization in these regions is most likely enhanced by irradiation of the cavity walls. 

\subitem (b) \textit{Potentially accretion-related features.}  These are thin, approximately linear features with high polarization fractions (usually 10--20\%) and extremely well ordered magnetic fields; however, these features do not appear to be associated with outflow cavities.  Examples include the sharp feature to the NE of \bhr IRS1; the magnetized filamentary structure around Ser-emb 8; the N--S equatorial-plane feature in B335; the ``arm-like'' structure in OMC3-MMS6; the ``bridge'' and streamers observed toward the two binary components in IRAS 16293; and one of the two narrow filaments to the south of Serpens SMM1-a.  We speculate that these features may be magnetized accretion streamers; however, this scenario has yet to be confirmed by kinematic observations of dense-gas tracers.

\end{enumerate}

With the advent of ALMA, our ability to probe the structure of magnetic fields in protostellar cores has vastly improved.  Upcoming surveys of large numbers of young, embedded sources will soon reveal how common is each of the scenarios that individual-source studies have recently unveiled: natal hourglass-shaped fields, magnetic fields affected by bipolar outflows, and possible magnetized accretion streamers.  Furthermore, observations of large-scale magnetic fields using current and upcoming single-dish polarimeters on instruments such as the Stratospheric Observatory for Infrared Astronomy (SOFIA; \citealt{Vaillancourt2007}), the James-Clerk-Maxell Telescope (JCMT; e.g., the BISTRO survey: \citealt{WardThompson2017}); the BLAST-TNG balloon-borne experiment \citep{Galitzki2014}, the IRAM 30\,m telescope \citep{Ritacco2017}, and the Large Millimeter Telescope (LMT) will allow us to understand how the larger-scale magnetic environment connects with the myriad small-scale magnetic field morphologies revealed by ALMA.

\acknowledgments
The authors thank the anonymous referee, whose insightful comments led to substantial improvements in the manuscript.
The authors acknowledge the support of Gerald Schieven at the North American ALMA Science Center, and Walker Lu and Toshinobu Takagi at the East Asian ALMA Regional Center.
C.L.H.H. and V.J.M.L.G. acknowledge helpful discussions with the ALMA Fellows, Ruud Visser, Lars Kristensen, Ana\"elle Maury, Zhi-Yun Li, Phil Myers, Ian Stephens, Sarah Sadavoy, and Felipe Alves. 
C.L.H.H. acknowledges the support of both the NAOJ Fellowship as well as JSPS KAKENHI grant 18K13586.
V.J.M.L.G. acknowledges the support of the ESO Studentship Program.
J.M.G. acknowledges the support of the Spanish MINECO AYA2017-84390-C2-R grant, and the Joint ALMA Observatory Visitor Program.
This paper makes use of the following ALMA data: ADS/JAO.ALMA\#2013.1.00518.S and ADS/JAO.ALMA\#2017.1.00655.S.
ALMA is a partnership of ESO (representing its member states), NSF (USA) and NINS (Japan), together with NRC (Canada), MOST and ASIAA (Taiwan), and KASI (Republic of Korea), in cooperation with the Republic of Chile. The Joint ALMA Observatory is operated by ESO, AUI/NRAO and NAOJ.
The National Radio Astronomy Observatory is a facility of the National Science Foundation operated under cooperative agreement by Associated Universities, Inc. 

\textit{Facilities:} ALMA.

\textit{Software:} APLpy, an open-source plotting package for Python hosted at \url{http://aplpy.github.com} \citep{Robitaille2012}.  CASA \citep{McMullin2007}.  Astropy \citep{Astropy2018}.

\bibliography{ms}

\begin{thebibliography}{}
\expandafter\ifx\csname natexlab\endcsname\relax\def\natexlab#1{#1}\fi

\bibitem[{{Adatia} \& {Rudge}(1975)}]{Adatia1975}
{Adatia}, N.~A., \& {Rudge}, A.~W. 1975, Electronics Letters, 11, 513

\bibitem[{{Aikawa} {et~al.}(2001){Aikawa}, {Ohashi}, {Inutsuka}, {Herbst}, \&
  {Takakuwa}}]{Aikawa2001}
{Aikawa}, Y., {Ohashi}, N., {Inutsuka}, S.-i., {Herbst}, E., \& {Takakuwa}, S.
  2001, \apj, 552, 639

\bibitem[{{Allen} {et~al.}(2003){Allen}, {Li}, \& {Shu}}]{Allen2003}
{Allen}, A., {Li}, Z.-Y., \& {Shu}, F.~H. 2003, \apj, 599, 363

\bibitem[{Alves {et~al.}(2019)Alves, Caselli, Girart, Segura-Cox, Franco,
  Schmiedeke, \& Zhao}]{Alves2019}
Alves, F.~O., Caselli, P., Girart, J.~M., {et~al.} 2019, Science, 366, 90

\bibitem[{{Alves} {et~al.}(2014){Alves}, {Frau}, {Girart}, {Franco}, {Santos},
  \& {Wiesemeyer}}]{Alves2014}
{Alves}, F.~O., {Frau}, P., {Girart}, J.~M., {et~al.} 2014, \aap, 569, L1

\bibitem[{{Anderl} {et~al.}(2016){Anderl}, {Maret}, {Cabrit}, {Belloche},
  {Maury}, {Andr{\'e}}, {Codella}, {Bacmann}, {Bontemps}, {Podio}, {Gueth}, \&
  {Bergin}}]{Anderl2016}
{Anderl}, S., {Maret}, S., {Cabrit}, S., {et~al.} 2016, \aap, 591, A3

\bibitem[{{Andersson} {et~al.}(2015){Andersson}, {Lazarian}, \&
  {Vaillancourt}}]{Andersson2015}
{Andersson}, B.-G., {Lazarian}, A., \& {Vaillancourt}, J.~E. 2015, \araa, 53,
  501

\bibitem[{{Astropy Collaboration} {et~al.}(2018){Astropy Collaboration},
  {Price-Whelan}, {Sip{\H o}cz}, {G{\"u}nther}, {Lim}, {Crawford}, {Conseil},
  {Shupe}, {Craig}, {Dencheva}, {Ginsburg}, {VanderPlas}, {Bradley},
  {P{\'e}rez-Su{\'a}rez}, {de Val-Borro}, {Aldcroft}, {Cruz}, {Robitaille},
  {Tollerud}, {Ardelean}, {Babej}, {Bach}, {Bachetti}, {Bakanov}, {Bamford},
  {Barentsen}, {Barmby}, {Baumbach}, {Berry}, {Biscani}, {Boquien}, {Bostroem},
  {Bouma}, {Brammer}, {Bray}, {Breytenbach}, {Buddelmeijer}, {Burke},
  {Calderone}, {Cano Rodr{\'{\i}}guez}, {Cara}, {Cardoso}, {Cheedella},
  {Copin}, {Corrales}, {Crichton}, {D'Avella}, {Deil}, {Depagne}, {Dietrich},
  {Donath}, {Droettboom}, {Earl}, {Erben}, {Fabbro}, {Ferreira}, {Finethy},
  {Fox}, {Garrison}, {Gibbons}, {Goldstein}, {Gommers}, {Greco}, {Greenfield},
  {Groener}, {Grollier}, {Hagen}, {Hirst}, {Homeier}, {Horton}, {Hosseinzadeh},
  {Hu}, {Hunkeler}, {Ivezi{\'c}}, {Jain}, {Jenness}, {Kanarek}, {Kendrew},
  {Kern}, {Kerzendorf}, {Khvalko}, {King}, {Kirkby}, {Kulkarni}, {Kumar},
  {Lee}, {Lenz}, {Littlefair}, {Ma}, {Macleod}, {Mastropietro}, {McCully},
  {Montagnac}, {Morris}, {Mueller}, {Mumford}, {Muna}, {Murphy}, {Nelson},
  {Nguyen}, {Ninan}, {N{\"o}the}, {Ogaz}, {Oh}, {Parejko}, {Parley}, {Pascual},
  {Patil}, {Patil}, {Plunkett}, {Prochaska}, {Rastogi}, {Reddy Janga},
  {Sabater}, {Sakurikar}, {Seifert}, {Sherbert}, {Sherwood-Taylor}, {Shih},
  {Sick}, {Silbiger}, {Singanamalla}, {Singer}, {Sladen}, {Sooley},
  {Sornarajah}, {Streicher}, {Teuben}, {Thomas}, {Tremblay}, {Turner},
  {Terr{\'o}n}, {van Kerkwijk}, {de la Vega}, {Watkins}, {Weaver}, {Whitmore},
  {Woillez}, {Zabalza}, \& {Astropy Contributors}}]{Astropy2018}
{Astropy Collaboration}, {Price-Whelan}, A.~M., {Sip{\H o}cz}, B.~M., {et~al.}
  2018, \aj, 156, 123

\bibitem[{{Bohlin} {et~al.}(1978){Bohlin}, {Savage}, \& {Drake}}]{Bohlin1978}
{Bohlin}, R.~C., {Savage}, B.~D., \& {Drake}, J.~F. 1978, \apj, 224, 132

\bibitem[{{Boss}(2000)}]{Boss2000}
{Boss}, A.~P. 2000, \apjl, 545, L61

\bibitem[{{Bourke}(2001)}]{Bourke2001}
{Bourke}, T.~L. 2001, \apjl, 554, L91

\bibitem[{{Bourke} {et~al.}(1997){Bourke}, {Garay}, {Lehtinen},
  {K{\"o}hnenkamp}, {Launhardt}, {Nyman}, {May}, {Robinson}, \&
  {Hyland}}]{Bourke1997}
{Bourke}, T.~L., {Garay}, G., {Lehtinen}, K.~K., {et~al.} 1997, \apj, 476, 781

\bibitem[{{Brogan} {et~al.}(2018){Brogan}, {Hunter}, \&
  {Fomalont}}]{Brogan2018}
{Brogan}, C.~L., {Hunter}, T.~R., \& {Fomalont}, E.~B. 2018, arXiv e-prints,
  arXiv:1805.05266

\bibitem[{{Chandler} {et~al.}(2005){Chandler}, {Brogan}, {Shirley}, \&
  {Loinard}}]{Chandler2005}
{Chandler}, C.~J., {Brogan}, C.~L., {Shirley}, Y.~L., \& {Loinard}, L. 2005,
  \apj, 632, 371

\bibitem[{{Chen} {et~al.}(2008){Chen}, {Launhardt}, {Bourke}, {Henning}, \&
  {Barnes}}]{Chen2008}
{Chen}, X., {Launhardt}, R., {Bourke}, T.~L., {Henning}, T., \& {Barnes}, P.~J.
  2008, \apj, 683, 862

\bibitem[{{Chiang} {et~al.}(2012){Chiang}, {Looney}, \& {Tobin}}]{Chiang2012}
{Chiang}, H.-F., {Looney}, L.~W., \& {Tobin}, J.~J. 2012, \apj, 756, 168

\bibitem[{{Christie} {et~al.}(2011){Christie}, {Viti}, {Williams}, {Girart}, \&
  {Morata}}]{Christie2011}
{Christie}, H., {Viti}, S., {Williams}, D.~A., {Girart}, J.~M., \& {Morata}, O.
  2011, \mnras, 416, 288

\bibitem[{{Chu} \& {Turrin}(1973)}]{Chu1973}
{Chu}, T.-S., \& {Turrin}, R. 1973, IEEE Transactions on Antennas and
  Propagation, 21, 339

\bibitem[{{Contopoulos} \& {Jappel}(1974)}]{IAU1974}
{Contopoulos}, G., \& {Jappel}, A., eds. 1974, Transactions of the IAU, Vol.
  XVB 1974, Proceedings of the Fifteenth General Assembly and Extraordinary
  General Assembly, Vol.~15, 165--167

\bibitem[{{Cortes} {et~al.}(2016){Cortes}, {Girart}, {Hull}, {Sridharan},
  {Louvet}, {Plambeck}, {Li}, {Crutcher}, \& {Lai}}]{Cortes2016}
{Cortes}, P.~C., {Girart}, J.~M., {Hull}, C.~L.~H., {et~al.} 2016, \apjl, 825,
  L15

\bibitem[{{Cortes} {et~al.}(2019){Cortes}, {Hull}, {Girart}, {Orquera-Rojas},
  {Sridharan}, {Li}, {Louvet}, {Cortes}, {Le Gouellec}, {Crutcher}, \&
  {Lai}}]{Cortes2019}
{Cortes}, P.~C., {Hull}, C.~L.~H., {Girart}, J.~M., {et~al.} 2019, \apj,
  submitted

\bibitem[{{Draine} \& {Weingartner}(1996)}]{Draine1996}
{Draine}, B.~T., \& {Weingartner}, J.~C. 1996, \apj, 470, 551

\bibitem[{{Drozdovskaya} {et~al.}(2015){Drozdovskaya}, {Walsh}, {Visser},
  {Harsono}, \& {van Dishoeck}}]{Drozdovskaya2015}
{Drozdovskaya}, M.~N., {Walsh}, C., {Visser}, R., {Harsono}, D., \& {van
  Dishoeck}, E.~F. 2015, \mnras, 451, 3836

\bibitem[{{Fiedler} \& {Mouschovias}(1993)}]{Fiedler1993}
{Fiedler}, R.~A., \& {Mouschovias}, T.~C. 1993, \apj, 415, 680

\bibitem[{{Fissel} {et~al.}(2019){Fissel}, {Ade}, {Angil{\`e}}, {Ashton},
  {Benton}, {Chen}, {Cunningham}, {Devlin}, {Dober}, {Friesen}, {Fukui},
  {Galitzki}, {Gandilo}, {Goodman}, {Green}, {Jones}, {Klein}, {King},
  {Korotkov}, {Li}, {Lowe}, {Martin}, {Matthews}, {Moncelsi}, {Nakamura},
  {Netterfield}, {Newmark}, {Novak}, {Pascale}, {Poidevin}, {Santos}, {Savini},
  {Scott}, {Shariff}, {Soler}, {Thomas}, {Tucker}, {Tucker}, {Ward-Thompson},
  \& {Zucker}}]{Fissel2019}
{Fissel}, L.~M., {Ade}, P. A.~R., {Angil{\`e}}, F.~E., {et~al.} 2019, \apj,
  878, 110

\bibitem[{{Frank} {et~al.}(2014){Frank}, {Ray}, {Cabrit}, {Hartigan}, {Arce},
  {Bacciotti}, {Bally}, {Benisty}, {Eisl{\"o}ffel}, {G{\"u}del}, {Lebedev},
  {Nisini}, \& {Raga}}]{Frank2014}
{Frank}, A., {Ray}, T.~P., {Cabrit}, S., {et~al.} 2014, in Protostars and
  Planets VI, ed. H.~{Beuther}, R.~S. {Klessen}, C.~P. {Dullemond}, \&
  T.~{Henning} (Tucson, Arizona: University of Arizona Press), 451--474

\bibitem[{{Frau} {et~al.}(2011){Frau}, {Galli}, \& {Girart}}]{Frau2011}
{Frau}, P., {Galli}, D., \& {Girart}, J.~M. 2011, \aap, 535, A44

\bibitem[{{Galametz} {et~al.}(2019){Galametz}, {Maury}, {Valdivia}, {Testi},
  {Belloche}, \& {Andre}}]{Galametz2019}
{Galametz}, M., {Maury}, A.~J., {Valdivia}, V., {et~al.} 2019, arXiv e-prints,
  arXiv:1910.04652

\bibitem[{{Galametz} {et~al.}(2018){Galametz}, {Maury}, {Girart}, {Rao},
  {Zhang}, {Gaudel}, {Valdivia}, {Keto}, \& {Lai}}]{Galametz2018}
{Galametz}, M., {Maury}, A., {Girart}, J.~M., {et~al.} 2018, \aap, 616, A139

\bibitem[{{Galitzki} {et~al.}(2014){Galitzki}, {Ade}, {Angil{\`e}}, {Ashton},
  {Beall}, {Becker}, {Bradford}, {Che}, {Cho}, {Devlin}, {Dober}, {Fissel},
  {Fukui}, {Gao}, {Groppi}, {Hillbrand}, {Hilton}, {Hubmayr}, {Irwin}, {Klein},
  {van Lanen}, {Li}, {Li}, {Lourie}, {Mani}, {Martin}, {Mauskopf}, {Nakamura},
  {Novak}, {Pappas}, {Pascale}, {Pisano}, {Santos}, {Savini}, {Scott},
  {Stanchfield}, {Tucker}, {Ullom}, {Underhill}, {Vissers}, \&
  {Ward-Thompson}}]{Galitzki2014}
{Galitzki}, N., {Ade}, P.~A.~R., {Angil{\`e}}, F.~E., {et~al.} 2014, Journal of
  Astronomical Instrumentation, 3, 1440001

\bibitem[{{Galli} \& {Shu}(1993{\natexlab{a}})}]{Galli1993a}
{Galli}, D., \& {Shu}, F.~H. 1993{\natexlab{a}}, \apj, 417, 220

\bibitem[{{Galli} \& {Shu}(1993{\natexlab{b}})}]{Galli1993b}
---. 1993{\natexlab{b}}, \apj, 417, 243

\bibitem[{{Garay} {et~al.}(1998){Garay}, {K{\"o}hnenkamp}, {Bourke},
  {Rodr{\'{\i}}guez}, \& {Lehtinen}}]{Garay1998}
{Garay}, G., {K{\"o}hnenkamp}, I., {Bourke}, T.~L., {Rodr{\'{\i}}guez}, L.~F.,
  \& {Lehtinen}, K.~K. 1998, \apj, 509, 768

\bibitem[{{Girart} {et~al.}(1999){Girart}, {Crutcher}, \& {Rao}}]{Girart1999}
{Girart}, J.~M., {Crutcher}, R.~M., \& {Rao}, R. 1999, \apjl, 525, L109

\bibitem[{{Girart} {et~al.}(2006){Girart}, {Rao}, \& {Marrone}}]{Girart2006}
{Girart}, J.~M., {Rao}, R., \& {Marrone}, D.~P. 2006, Science, 313, 812

\bibitem[{{Girart} {et~al.}(2005){Girart}, {Viti}, {Estalella}, \&
  {Williams}}]{Girart2005}
{Girart}, J.~M., {Viti}, S., {Estalella}, R., \& {Williams}, D.~A. 2005, \aap,
  439, 601

\bibitem[{{Girart} {et~al.}(2002){Girart}, {Viti}, {Williams}, {Estalella}, \&
  {Ho}}]{Girart2002}
{Girart}, J.~M., {Viti}, S., {Williams}, D.~A., {Estalella}, R., \& {Ho},
  P.~T.~P. 2002, \aap, 388, 1004

\bibitem[{{Girart} {et~al.}(1994){Girart}, {Rodriiguez}, {Anglada},
  {Estalella}, {Torrelles}, {Marti}, {Pena}, {Ayala}, {Curiel}, \&
  {Noriega-Crespo}}]{Girart1994}
{Girart}, J.~M., {Rodriiguez}, L.~F., {Anglada}, G., {et~al.} 1994, \apj, 435,
  L145

\bibitem[{{Gold}(1952)}]{Gold1952}
{Gold}, T. 1952, \mnras, 112, 215

\bibitem[{{Gusdorf} {et~al.}(2015){Gusdorf}, {Riquelme}, {Anderl},
  {Eisl{\"o}ffel}, {Codella}, {G{\'o}mez-Ruiz}, {Graf}, {Kristensen},
  {Leurini}, {Parise}, {Requena-Torres}, {Ricken}, \&
  {G{\"u}sten}}]{Gusdorf2015}
{Gusdorf}, A., {Riquelme}, D., {Anderl}, S., {et~al.} 2015, \aap, 575, A98

\bibitem[{{Hennebelle} {et~al.}(2011){Hennebelle}, {Commer{\c{c}}on}, {Joos},
  {Klessen}, {Krumholz}, {Tan}, \& {Teyssier}}]{Hennebelle2011}
{Hennebelle}, P., {Commer{\c{c}}on}, B., {Joos}, M., {et~al.} 2011, \aap, 528,
  A72

\bibitem[{{Herbst}(1982)}]{Herbst1982}
{Herbst}, E. 1982, \aap, 111, 76

\bibitem[{{Hoang} {et~al.}(2018){Hoang}, {Cho}, \& {Lazarian}}]{Hoang2018}
{Hoang}, T., {Cho}, J., \& {Lazarian}, A. 2018, \apj, 852, 129

\bibitem[{{Hoang} \& {Lazarian}(2009)}]{Hoang2009}
{Hoang}, T., \& {Lazarian}, A. 2009, \apj, 697, 1316

\bibitem[{{Hull} \& {Plambeck}(2015)}]{Hull2015b}
{Hull}, C.~L.~H., \& {Plambeck}, R.~L. 2015, Journal of Astronomical
  Instrumentation, 4, 1550005

\bibitem[{Hull \& Zhang(2019)}]{HullZhang2019}
Hull, C. L.~H., \& Zhang, Q. 2019, Frontiers in Astronomy and Space Sciences,
  6, 3

\bibitem[{{Hull} {et~al.}(2013){Hull}, {Plambeck}, {Bolatto}, {Bower},
  {Carpenter}, {Crutcher}, {Fiege}, {Franzmann}, {Hakobian}, {Heiles}, {Houde},
  {Hughes}, {Jameson}, {Kwon}, {Lamb}, {Looney}, {Matthews}, {Mundy}, {Pillai},
  {Pound}, {Stephens}, {Tobin}, {Vaillancourt}, {Volgenau}, \&
  {Wright}}]{Hull2013}
{Hull}, C.~L.~H., {Plambeck}, R.~L., {Bolatto}, A.~D., {et~al.} 2013, \apj,
  768, 159

\bibitem[{{Hull} {et~al.}(2014){Hull}, {Plambeck}, {Kwon}, {Bower},
  {Carpenter}, {Crutcher}, {Fiege}, {Franzmann}, {Hakobian}, {Heiles}, {Houde},
  {Hughes}, {Lamb}, {Looney}, {Marrone}, {Matthews}, {Pillai}, {Pound},
  {Rahman}, {Sandell}, {Stephens}, {Tobin}, {Vaillancourt}, {Volgenau}, \&
  {Wright}}]{Hull2014}
{Hull}, C.~L.~H., {Plambeck}, R.~L., {Kwon}, W., {et~al.} 2014, \apjs, 213, 13

\bibitem[{{Hull} {et~al.}(2017{\natexlab{a}}){Hull}, {Girart}, {Tychoniec},
  {Rao}, {Cort{\'e}s}, {Pokhrel}, {Zhang}, {Houde}, {Dunham}, {Kristensen},
  {Lai}, {Li}, \& {Plambeck}}]{Hull2017b}
{Hull}, C.~L.~H., {Girart}, J.~M., {Tychoniec}, {\L}., {et~al.}
  2017{\natexlab{a}}, \apj, 847, 92

\bibitem[{{Hull} {et~al.}(2017{\natexlab{b}}){Hull}, {Mocz}, {Burkhart},
  {Goodman}, {Girart}, {Cort{\'e}s}, {Hernquist}, {Springel}, {Li}, \&
  {Lai}}]{Hull2017a}
{Hull}, C.~L.~H., {Mocz}, P., {Burkhart}, B., {et~al.} 2017{\natexlab{b}},
  \apjl, 842, L9

\bibitem[{{Hull} {et~al.}(2018){Hull}, {Yang}, {Li}, {Kataoka}, {Stephens},
  {Andrews}, {Bai}, {Cleeves}, {Hughes}, {Looney}, {P{\'e}rez}, \&
  {Wilner}}]{Hull2018a}
{Hull}, C.~L.~H., {Yang}, H., {Li}, Z.-Y., {et~al.} 2018, \apj, 860, 82

\bibitem[{{Imai} {et~al.}(2016){Imai}, {Sakai}, {Oya}, {L{\'o}pez-Sepulcre},
  {Watanabe}, {Ceccarelli}, {Lefloch}, {Caux}, {Vastel}, {Kahane}, {Sakai},
  {Hirota}, {Aikawa}, \& {Yamamoto}}]{Imai2016}
{Imai}, M., {Sakai}, N., {Oya}, Y., {et~al.} 2016, \apj, 830, L37

\bibitem[{{Jensen} \& {Akeson}(2014)}]{JensenAkeson2014}
{Jensen}, E.~L.~N., \& {Akeson}, R. 2014, \nat, 511, 567

\bibitem[{{J{\o}rgensen}(2004)}]{Jorgensen2004b}
{J{\o}rgensen}, J.~K. 2004, \aap, 424, 589

\bibitem[{{J{\o}rgensen} {et~al.}(2007){J{\o}rgensen}, {Bourke}, {Myers}, {Di
  Francesco}, {van Dishoeck}, {Lee}, {Ohashi}, {Sch{\"o}ier}, {Takakuwa},
  {Wilner}, \& {Zhang}}]{Jorgensen2007}
{J{\o}rgensen}, J.~K., {Bourke}, T.~L., {Myers}, P.~C., {et~al.} 2007, \apj,
  659, 479

\bibitem[{{Karska} {et~al.}(2018){Karska}, {Kaufman}, {Kristensen}, {van
  Dishoeck}, {Herczeg}, {Mottram}, {Tychoniec}, {Lindberg}, {Evans}, {Green},
  {Yang}, {Gusdorf}, {Itrich}, \& {Si{\'o}dmiak}}]{Karska2018}
{Karska}, A., {Kaufman}, M.~J., {Kristensen}, L.~E., {et~al.} 2018, \apjs, 235,
  30

\bibitem[{{Kauffmann} {et~al.}(2008){Kauffmann}, {Bertoldi}, {Bourke}, {Evans},
  \& {Lee}}]{Kauffmann2008}
{Kauffmann}, J., {Bertoldi}, F., {Bourke}, T.~L., {Evans}, II, N.~J., \& {Lee},
  C.~W. 2008, \aap, 487, 993

\bibitem[{{Koch} {et~al.}(2012){Koch}, {Tang}, \& {Ho}}]{Koch2012}
{Koch}, P.~M., {Tang}, Y.-W., \& {Ho}, P.~T.~P. 2012, \apj, 747, 79

\bibitem[{{Koch} {et~al.}(2018){Koch}, {Tang}, {Ho}, {Yen}, {Su}, \&
  {Takakuwa}}]{Koch2018}
{Koch}, P.~M., {Tang}, Y.-W., {Ho}, P.~T.~P., {et~al.} 2018, \apj, 855, 39

\bibitem[{{Kratter} {et~al.}(2010){Kratter}, {Matzner}, {Krumholz}, \&
  {Klein}}]{Kratter2010}
{Kratter}, K.~M., {Matzner}, C.~D., {Krumholz}, M.~R., \& {Klein}, R.~I. 2010,
  \apj, 708, 1585

\bibitem[{{Kristensen} {et~al.}(2017){Kristensen}, {van Dishoeck}, {Mottram},
  {Karska}, {Y{\i}ld{\i}z}, {Bergin}, {Bjerkeli}, {Cabrit}, {Doty}, {Evans},
  {Gusdorf}, {Harsono}, {Herczeg}, {Johnstone}, {J{\o}rgensen}, {van Kempen},
  {Lee}, {Maret}, {Tafalla}, {Visser}, \& {Wampfler}}]{Kristensen2017}
{Kristensen}, L.~E., {van Dishoeck}, E.~F., {Mottram}, J.~C., {et~al.} 2017,
  \aap, 605, A93

\bibitem[{{Krumholz} \& {Federrath}(2019)}]{KrumholzFederrath2019}
{Krumholz}, M.~R., \& {Federrath}, C. 2019, Frontiers in Astronomy and Space
  Sciences, 6, 7

\bibitem[{{Kwon} {et~al.}(2015){Kwon}, {Fern{\'a}ndez-L{\'o}pez}, {Stephens},
  \& {Looney}}]{Kwon2015}
{Kwon}, W., {Fern{\'a}ndez-L{\'o}pez}, M., {Stephens}, I.~W., \& {Looney},
  L.~W. 2015, \apj, 814, 43

\bibitem[{{Kwon} {et~al.}(2009){Kwon}, {Looney}, {Mundy}, {Chiang}, \&
  {Kemball}}]{Kwon2009}
{Kwon}, W., {Looney}, L.~W., {Mundy}, L.~G., {Chiang}, H.-F., \& {Kemball},
  A.~J. 2009, \apj, 696, 841

\bibitem[{{Kwon} {et~al.}(2019){Kwon}, {Stephens}, {Tobin}, {Looney}, {Li},
  {van der Tak}, \& {Crutcher}}]{Kwon2019}
{Kwon}, W., {Stephens}, I.~W., {Tobin}, J.~J., {et~al.} 2019, \apj, 879, 25

\bibitem[{{Lazarian} \& {Hoang}(2007{\natexlab{a}})}]{LazarianHoang2007a}
{Lazarian}, A., \& {Hoang}, T. 2007{\natexlab{a}}, \mnras, 378, 910

\bibitem[{{Lazarian} \& {Hoang}(2007{\natexlab{b}})}]{LazarianHoang2007b}
---. 2007{\natexlab{b}}, \apj, 669, L77

\bibitem[{{Le Gouellec} {et~al.}(2019){Le Gouellec}, {Hull}, {Maury}, {Girart},
  {Tychoniec}, {Kristensen}, {Li}, {Louvet}, {Cortes}, \&
  {Rao}}]{LeGouellec2019a}
{Le Gouellec}, V. J.~M., {Hull}, C. L.~H., {Maury}, A.~J., {et~al.} 2019, arXiv
  e-prints, arXiv:1909.00046

\bibitem[{{Lee} {et~al.}(2017){Lee}, {Lee}, {Dunham}, {Tatematsu}, {Choi},
  {Bergin}, \& {Evans}}]{JELee2017}
{Lee}, J.-E., {Lee}, S., {Dunham}, M.~M., {et~al.} 2017, Nature Astronomy, 1,
  0172

\bibitem[{{Lee} {et~al.}(2016){Lee}, {Dunham}, {Myers}, {Arce}, {Bourke},
  {Goodman}, {J{\o}rgensen}, {Kristensen}, {Offner}, {Pineda}, {Tobin}, \&
  {Vorobyov}}]{Lee2016}
{Lee}, K.~I., {Dunham}, M.~M., {Myers}, P.~C., {et~al.} 2016, \apjl, 820, L2

\bibitem[{{Li} {et~al.}(2017){Li}, {Liu}, {Hasegawa}, \& {Hirano}}]{LiJ2017}
{Li}, J. I.-H., {Liu}, H.~B., {Hasegawa}, Y., \& {Hirano}, N. 2017, \apj, 840,
  72

\bibitem[{{Looney} {et~al.}(2000){Looney}, {Mundy}, \& {Welch}}]{Looney2000}
{Looney}, L.~W., {Mundy}, L.~G., \& {Welch}, W.~J. 2000, \apj, 529, 477

\bibitem[{{Maury} {et~al.}(2018){Maury}, {Girart}, {Zhang}, {Hennebelle},
  {Keto}, {Rao}, {Lai}, {Ohashi}, \& {Galametz}}]{Maury2018}
{Maury}, A.~J., {Girart}, J.~M., {Zhang}, Q., {et~al.} 2018, \mnras, 477, 2760

\bibitem[{{McKee} \& {Ostriker}(2007)}]{McKee2007}
{McKee}, C.~F., \& {Ostriker}, E.~C. 2007, \araa, 45, 565

\bibitem[{{McKee} {et~al.}(1993){McKee}, {Zweibel}, {Goodman}, \&
  {Heiles}}]{McKee1993}
{McKee}, C.~F., {Zweibel}, E.~G., {Goodman}, A.~A., \& {Heiles}, C. 1993, in
  Protostars and Planets III, ed. E.~H. {Levy} \& J.~I. {Lunine} (Tucson,
  Arizona: University of Arizona Press), 327

\bibitem[{{McMullin} {et~al.}(2007){McMullin}, {Waters}, {Schiebel}, {Young},
  \& {Golap}}]{McMullin2007}
{McMullin}, J.~P., {Waters}, B., {Schiebel}, D., {Young}, W., \& {Golap}, K.
  2007, in Astronomical Society of the Pacific Conference Series, Vol. 376,
  Astronomical Data Analysis Software and Systems XVI, ed. R.~A. {Shaw},
  F.~{Hill}, \& D.~J. {Bell}, 127

\bibitem[{{Mestel} \& {Spitzer}(1956)}]{Mestel1956}
{Mestel}, L., \& {Spitzer}, Jr., L. 1956, \mnras, 116, 503

\bibitem[{{Miotello} {et~al.}(2014){Miotello}, {Testi}, {Lodato}, {Ricci},
  {Rosotti}, {Brooks}, {Maury}, \& {Natta}}]{Miotello2014}
{Miotello}, A., {Testi}, L., {Lodato}, G., {et~al.} 2014, \aap, 567, A32

\bibitem[{{Nagai} {et~al.}(2016){Nagai}, {Nakanishi}, {Paladino}, {Hull},
  {Cortes}, {Moellenbrock}, {Fomalont}, {Asada}, \& {Hada}}]{Nagai2016}
{Nagai}, H., {Nakanishi}, K., {Paladino}, R., {et~al.} 2016, \apj, 824, 132

\bibitem[{{Offner} \& {Chaban}(2017)}]{Offner2017}
{Offner}, S.~S.~R., \& {Chaban}, J. 2017, \apj, 847, 104

\bibitem[{{Offner} {et~al.}(2016){Offner}, {Dunham}, {Lee}, {Arce}, \&
  {Fielding}}]{Offner2016}
{Offner}, S.~S.~R., {Dunham}, M.~M., {Lee}, K.~I., {Arce}, H.~G., \&
  {Fielding}, D.~B. 2016, \apjl, 827, L11

\bibitem[{{Ossenkopf} \& {Henning}(1994)}]{Ossenkopf1994}
{Ossenkopf}, V., \& {Henning}, T. 1994, \aap, 291, 943

\bibitem[{{Parise} {et~al.}(2006){Parise}, {Belloche}, {Leurini}, {Schilke},
  {Wyrowski}, \& {G{\"u}sten}}]{Parise2006}
{Parise}, B., {Belloche}, A., {Leurini}, S., {et~al.} 2006, \aap, 454, L79

\bibitem[{{P{\'e}rez} {et~al.}(2012){P{\'e}rez}, {Carpenter}, {Chandler},
  {Isella}, {Andrews}, {Ricci}, {Calvet}, {Corder}, {Deller}, {Dullemond},
  {Greaves}, {Harris}, {Henning}, {Kwon}, {Lazio}, {Linz}, {Mundy}, {Sargent},
  {Storm}, {Testi}, \& {Wilner}}]{Perez2012}
{P{\'e}rez}, L.~M., {Carpenter}, J.~M., {Chandler}, C.~J., {et~al.} 2012,
  \apjl, 760, L17

\bibitem[{{Pineda} {et~al.}(2015){Pineda}, {Offner}, {Parker}, {Arce},
  {Goodman}, {Caselli}, {Fuller}, {Bourke}, \& {Corder}}]{Pineda2015}
{Pineda}, J.~E., {Offner}, S.~S.~R., {Parker}, R.~J., {et~al.} 2015, \nat, 518,
  213

\bibitem[{{Planck Collaboration} {et~al.}(2016{\natexlab{a}}){Planck
  Collaboration}, {Adam}, {Ade}, {Aghanim}, {Alves}, \& et~al.}]{PlanckXXXII}
{Planck Collaboration}, {Adam}, R., {Ade}, P.~A.~R., {et~al.}
  2016{\natexlab{a}}, \aap, 586, A135

\bibitem[{{Planck Collaboration} {et~al.}(2016{\natexlab{b}}){Planck
  Collaboration}, {Ade}, {Aghanim}, {Alves}, {Arnaud}, {Arzoumanian},
  {Ashdown}, {Aumont}, {Baccigalupi}, {Banday}, {Barreiro}, {Bartolo},
  {Battaner}, {Benabed}, {Beno{\^\i}t}, {Benoit-L{\'e}vy}, {Bernard},
  {Bersanelli}, {Bielewicz}, {Bock}, {Bonavera}, {Bond}, {Borrill}, {Bouchet},
  {Boulanger}, {Bracco}, {Burigana}, {Calabrese}, {Cardoso}, {Catalano},
  {Chiang}, {Christensen}, {Colombo}, {Combet}, {Couchot}, {Crill}, {Curto},
  {Cuttaia}, {Danese}, {Davies}, {Davis}, {de Bernardis}, {de Rosa}, {de
  Zotti}, {Delabrouille}, {Dickinson}, {Diego}, {Dole}, {Donzelli}, {Dor{\'e}},
  {Douspis}, {Ducout}, {Dupac}, {Efstathiou}, {Elsner}, {En{\ss}lin},
  {Eriksen}, {Falceta-Gon{\c c}alves}, {Falgarone}, {Ferri{\`e}re}, {Finelli},
  {Forni}, {Frailis}, {Fraisse}, {Franceschi}, {Frejsel}, {Galeotta}, {Galli},
  {Ganga}, {Ghosh}, {Giard}, {Gjerl{\o}w}, {Gonz{\'a}lez-Nuevo}, {G{\'o}rski},
  {Gregorio}, {Gruppuso}, {Gudmundsson}, {Guillet}, {Harrison}, {Helou},
  {Hennebelle}, {Henrot-Versill{\'e}}, {Hern{\'a}ndez-Monteagudo}, {Herranz},
  {Hildebrandt}, {Hivon}, {Holmes}, {Hornstrup}, {Huffenberger}, {Hurier},
  {Jaffe}, {Jaffe}, {Jones}, {Juvela}, {Keih{\"a}nen}, {Keskitalo}, {Kisner},
  {Knoche}, {Kunz}, {Kurki-Suonio}, {Lagache}, {Lamarre}, {Lasenby},
  {Lattanzi}, {Lawrence}, {Leonardi}, {Levrier}, {Liguori}, {Lilje},
  {Linden-V{\o}rnle}, {L{\'o}pez-Caniego}, {Lubin}, {Mac{\'{\i}}as-P{\'e}rez},
  {Maino}, {Mandolesi}, {Mangilli}, {Maris}, {Martin},
  {Mart{\'{\i}}nez-Gonz{\'a}lez}, {Masi}, {Matarrese}, {Melchiorri}, {Mendes},
  {Mennella}, {Migliaccio}, {Miville-Desch{\^e}nes}, {Moneti}, {Montier},
  {Morgante}, {Mortlock}, {Munshi}, {Murphy}, {Naselsky}, {Nati},
  {Netterfield}, {Noviello}, {Novikov}, {Novikov}, {Oppermann}, {Oxborrow},
  {Pagano}, {Pajot}, {Paladini}, {Paoletti}, {Pasian}, {Perotto}, {Pettorino},
  {Piacentini}, {Piat}, {Pierpaoli}, {Pietrobon}, {Plaszczynski},
  {Pointecouteau}, {Polenta}, {Ponthieu}, {Pratt}, {Prunet}, {Puget}, {Rachen},
  {Reinecke}, {Remazeilles}, {Renault}, {Renzi}, {Ristorcelli}, {Rocha},
  {Rossetti}, {Roudier}, {Rubi{\~n}o-Mart{\'{\i}}n}, {Rusholme}, {Sandri},
  {Santos}, {Savelainen}, {Savini}, {Scott}, {Soler}, {Stolyarov}, {Sudiwala},
  {Sutton}, {Suur-Uski}, {Sygnet}, {Tauber}, {Terenzi}, {Toffolatti}, {Tomasi},
  {Tristram}, {Tucci}, {Umana}, {Valenziano}, {Valiviita}, {Van Tent},
  {Vielva}, {Villa}, {Wade}, {Wandelt}, {Wehus}, {Ysard}, {Yvon}, \&
  {Zonca}}]{PlanckXXXV}
{Planck Collaboration}, {Ade}, P.~A.~R., {Aghanim}, N., {et~al.}
  2016{\natexlab{b}}, \aap, 586, A138

\bibitem[{{Rao} {et~al.}(2009){Rao}, {Girart}, {Marrone}, {Lai}, \&
  {Schnee}}]{Rao2009}
{Rao}, R., {Girart}, J.~M., {Marrone}, D.~P., {Lai}, S.-P., \& {Schnee}, S.
  2009, \apj, 707, 921

\bibitem[{{Reissl} {et~al.}(2016){Reissl}, {Wolf}, \& {Brauer}}]{Reissl2016}
{Reissl}, S., {Wolf}, S., \& {Brauer}, R. 2016, \aap, 593, A87

\bibitem[{{Ritacco} {et~al.}(2017){Ritacco}, {Ponthieu}, {Catalano}, {Adam},
  {Ade}, {Andr{\'e}}, {Beelen}, {Beno{\^\i}t}, {Bideaud}, {Billot}, {Bourrion},
  {Calvo}, {Coiffard}, {Comis}, {D{\'e}sert}, {Doyle}, {Goupy}, {Kramer},
  {Leclercq}, {Mac{\'{\i}}as-P{\'e}rez}, {Mauskopf}, {Maury}, {Mayet},
  {Monfardini}, {Pajot}, {Pascale}, {Perotto}, {Pisano}, {Rebolo-Iglesias},
  {Rev{\'e}ret}, {Rodriguez}, {Romero}, {Ruppin}, {Savini}, {Schuster},
  {Sievers}, {Thum}, {Triqueneaux}, {Tucker}, \& {Zylka}}]{Ritacco2017}
{Ritacco}, A., {Ponthieu}, N., {Catalano}, A., {et~al.} 2017, \aap, 599, A34

\bibitem[{{Robitaille} \& {Bressert}(2012)}]{Robitaille2012}
{Robitaille}, T., \& {Bressert}, E. 2012, {APLpy: Astronomical Plotting Library
  in Python}, Astrophysics Source Code Library, ascl:1208.017

\bibitem[{{Rudge} \& {Adatia}(1978)}]{Rudge1978}
{Rudge}, A.~W., \& {Adatia}, N.~A. 1978, IEEE Proceedings, 66, 1592

\bibitem[{{Sadavoy} {et~al.}(2018){Sadavoy}, {Myers}, {Stephens}, {Tobin},
  {Kwon}, {Segura-Cox}, {Henning}, {Commer{\c c}on}, \&
  {Looney}}]{Sadavoy2018c}
{Sadavoy}, S.~I., {Myers}, P.~C., {Stephens}, I.~W., {et~al.} 2018, \apj, 869,
  115

\bibitem[{{Segura-Cox} {et~al.}(2018){Segura-Cox}, {Looney}, {Tobin}, {Li},
  {Harris}, {Sadavoy}, {Dunham}, {Chandler}, {Kratter}, {P{\'e}rez}, \&
  {Melis}}]{SeguraCox2018}
{Segura-Cox}, D.~M., {Looney}, L.~W., {Tobin}, J.~J., {et~al.} 2018, \apj, 866,
  161

\bibitem[{{Seidensticker} \& {Schmidt-Kaler}(1989)}]{Seidensticker1989}
{Seidensticker}, K.~J., \& {Schmidt-Kaler}, T. 1989, \aap, 225, 192

\bibitem[{{Shu} {et~al.}(1987){Shu}, {Adams}, \& {Lizano}}]{Shu1987}
{Shu}, F.~H., {Adams}, F.~C., \& {Lizano}, S. 1987, \araa, 25, 23

\bibitem[{{Soler} {et~al.}(2013){Soler}, {Hennebelle}, {Martin},
  {Miville-Desch{\^e}nes}, {Netterfield}, \& {Fissel}}]{Soler2013}
{Soler}, J.~D., {Hennebelle}, P., {Martin}, P.~G., {et~al.} 2013, \apj, 774,
  128

\bibitem[{{Soler} {et~al.}(2017){Soler}, {Ade}, {Angil{\`e}}, {Ashton},
  {Benton}, {Devlin}, {Dober}, {Fissel}, {Fukui}, {Galitzki}, {Gand ilo},
  {Hennebelle}, {Klein}, {Li}, {Korotkov}, {Martin}, {Matthews}, {Moncelsi},
  {Netterfield}, {Novak}, {Pascale}, {Poidevin}, {Santos}, {Savini}, {Scott},
  {Shariff}, {Thomas}, {Tucker}, {Tucker}, \& {Ward-Thompson}}]{Soler2017}
{Soler}, J.~D., {Ade}, P.~A.~R., {Angil{\`e}}, F.~E., {et~al.} 2017, \aap, 603,
  A64

\bibitem[{{Soler, J.~D.}(2019)}]{Soler2019}
{Soler, J.~D.} 2019, A\&A, 629, A96

\bibitem[{{Spaans} {et~al.}(1995){Spaans}, {Hogerheijde}, {Mundy}, \& {van
  Dishoeck}}]{Spaans1995}
{Spaans}, M., {Hogerheijde}, M.~R., {Mundy}, L.~G., \& {van Dishoeck}, E.~F.
  1995, \apj, 455, L167

\bibitem[{{Stephens} {et~al.}(2013){Stephens}, {Looney}, {Kwon}, {Hull},
  {Plambeck}, {Crutcher}, {Chapman}, {Novak}, {Davidson}, {Vaillancourt},
  {Shinnaga}, \& {Matthews}}]{Stephens2013}
{Stephens}, I.~W., {Looney}, L.~W., {Kwon}, W., {et~al.} 2013, \apjl, 769, L15

\bibitem[{{Stephens} {et~al.}(2017{\natexlab{a}}){Stephens}, {Dunham}, {Myers},
  {Pokhrel}, {Sadavoy}, {Vorobyov}, {Tobin}, {Pineda}, {Offner}, {Lee},
  {Kristensen}, {J{\o}rgensen}, {Goodman}, {Bourke}, {Arce}, \&
  {Plunkett}}]{Stephens2017a}
{Stephens}, I.~W., {Dunham}, M.~M., {Myers}, P.~C., {et~al.}
  2017{\natexlab{a}}, \apj, 846, 16

\bibitem[{{Stephens} {et~al.}(2017{\natexlab{b}}){Stephens}, {Yang}, {Li},
  {Looney}, {Kataoka}, {Kwon}, {Fern{\'a}ndez-L{\'o}pez}, {Hull}, {Hughes},
  {Segura-Cox}, {Mundy}, {Crutcher}, \& {Rao}}]{Stephens2017b}
{Stephens}, I.~W., {Yang}, H., {Li}, Z.-Y., {et~al.} 2017{\natexlab{b}}, \apj,
  851, 55

\bibitem[{{Takahashi} {et~al.}(2019){Takahashi}, {Machida}, {Tomisaka}, {Ho},
  {Fomalont}, {Nakanishi}, \& {Girart}}]{Takahashi2019}
{Takahashi}, S., {Machida}, M.~N., {Tomisaka}, K., {et~al.} 2019, \apj, 872, 70

\bibitem[{{Taylor} \& {Williams}(1996)}]{Taylor1996}
{Taylor}, S.~D., \& {Williams}, D.~A. 1996, \mnras, 282, 1343

\bibitem[{{Testi} {et~al.}(2014){Testi}, {Birnstiel}, {Ricci}, {Andrews},
  {Blum}, {Carpenter}, {Dominik}, {Isella}, {Natta}, {Williams}, \&
  {Wilner}}]{Testi2014}
{Testi}, L., {Birnstiel}, T., {Ricci}, L., {et~al.} 2014, Protostars and
  Planets VI, 339

\bibitem[{{Thompson} {et~al.}(2017){Thompson}, {Moran}, \& {Swenson}}]{TMS}
{Thompson}, A.~R., {Moran}, J.~M., \& {Swenson}, George~W., J. 2017,
  {Interferometry and Synthesis in Radio Astronomy, 3rd Edition} (Springer),
  doi:10.1007/978-3-319-44431-4

\bibitem[{{Tobin} {et~al.}(2013{\natexlab{a}}){Tobin}, {Bergin}, {Hartmann},
  {Lee}, {Maret}, {Myers}, {Looney}, {Chiang}, \& {Friesen}}]{Tobin2013a}
{Tobin}, J.~J., {Bergin}, E.~A., {Hartmann}, L., {et~al.} 2013{\natexlab{a}},
  \apj, 765, 18

\bibitem[{{Tobin} {et~al.}(2013{\natexlab{b}}){Tobin}, {Chandler}, {Wilner},
  {Looney}, {Loinard}, {Chiang}, {Hartmann}, {Calvet}, {D'Alessio}, {Bourke},
  \& {Kwon}}]{Tobin2013c}
{Tobin}, J.~J., {Chandler}, C.~J., {Wilner}, D.~J., {et~al.}
  2013{\natexlab{b}}, \apj, 779, 93

\bibitem[{{Tobin} {et~al.}(2015){Tobin}, {Dunham}, {Looney}, {Li}, {Chandler},
  {Segura-Cox}, {Sadavoy}, {Melis}, {Harris}, {Perez}, {Kratter},
  {J{\o}rgensen}, {Plunkett}, \& {Hull}}]{Tobin2015a}
{Tobin}, J.~J., {Dunham}, M.~M., {Looney}, L.~W., {et~al.} 2015, \apj, 798, 61

\bibitem[{{Tobin} {et~al.}(2016{\natexlab{a}}){Tobin}, {Kratter}, {Persson},
  {Looney}, {Dunham}, {Segura-Cox}, {Li}, {Chandler}, {Sadavoy}, {Harris},
  {Melis}, \& {P{\'e}rez}}]{Tobin2016b}
{Tobin}, J.~J., {Kratter}, K.~M., {Persson}, M.~V., {et~al.}
  2016{\natexlab{a}}, \nat, 538, 483

\bibitem[{{Tobin} {et~al.}(2016{\natexlab{b}}){Tobin}, {Looney}, {Li},
  {Chandler}, {Dunham}, {Segura-Cox}, {Sadavoy}, {Melis}, {Harris}, {Kratter},
  \& {Perez}}]{Tobin2016a}
{Tobin}, J.~J., {Looney}, L.~W., {Li}, Z.-Y., {et~al.} 2016{\natexlab{b}},
  \apj, 818, 73

\bibitem[{{Tobin} {et~al.}(2018){Tobin}, {Looney}, {Li}, {Sadavoy}, {Dunham},
  {Segura-Cox}, {Kratter}, {Chandler}, {Melis}, {Harris}, \&
  {Perez}}]{Tobin2018}
---. 2018, \apj, 867, 43

\bibitem[{{Tobin} {et~al.}(2019){Tobin}, {Bourke}, {Mader}, {Kristensen},
  {Arce}, {Gueth}, {Gusdorf}, {Codella}, {Leurini}, \& {Chen}}]{Tobin2019}
{Tobin}, J.~J., {Bourke}, T.~L., {Mader}, S., {et~al.} 2019, \apj, 870, 81

\bibitem[{{Vaillancourt}(2006)}]{Vaillancourt2006}
{Vaillancourt}, J.~E. 2006, \pasp, 118, 1340

\bibitem[{{Vaillancourt} {et~al.}(2007){Vaillancourt}, {Chuss}, {Crutcher},
  {Dotson}, {Dowell}, {Harper}, {Hildebrand}, {Jones}, {Lazarian}, {Novak}, \&
  {Werner}}]{Vaillancourt2007}
{Vaillancourt}, J.~E., {Chuss}, D.~T., {Crutcher}, R.~M., {et~al.} 2007, in
  \procspie, Vol. 6678, Infrared Spaceborne Remote Sensing and Instrumentation
  XV, 66780D

\bibitem[{{Valdivia} {et~al.}(2019){Valdivia}, {Maury}, {Brauer}, {Hennebelle},
  {Galametz}, {Guillet}, \& {Reissl}}]{Valdivia2019}
{Valdivia}, V., {Maury}, A., {Brauer}, R., {et~al.} 2019, \mnras, 488, 4897

\bibitem[{{van Kempen} {et~al.}(2009{\natexlab{a}}){van Kempen}, {van
  Dishoeck}, {G{\"u}sten}, {Kristensen}, {Schilke}, {Hogerheijde}, {Boland},
  {Nefs}, {Menten}, {Baryshev}, \& {Wyrowski}}]{vanKempen2009a}
{van Kempen}, T.~A., {van Dishoeck}, E.~F., {G{\"u}sten}, R., {et~al.}
  2009{\natexlab{a}}, \aap, 501, 633

\bibitem[{{van Kempen} {et~al.}(2009{\natexlab{b}}){van Kempen}, {van
  Dishoeck}, {G{\"u}sten}, {Kristensen}, {Schilke}, {Hogerheijde}, {Boland},
  {Menten}, \& {Wyrowski}}]{vanKempen2009c}
---. 2009{\natexlab{b}}, \aap, 507, 1425

\bibitem[{{Vasyunina} {et~al.}(2012){Vasyunina}, {Vasyunin}, {Herbst}, \&
  {Linz}}]{Vasyunina2012}
{Vasyunina}, T., {Vasyunin}, A.~I., {Herbst}, E., \& {Linz}, H. 2012, \apj,
  751, 105

\bibitem[{{Visser} {et~al.}(2012){Visser}, {Kristensen}, {Bruderer}, {van
  Dishoeck}, {Herczeg}, {Brinch}, {Doty}, {Harsono}, \& {Wolfire}}]{Visser2012}
{Visser}, R., {Kristensen}, L.~E., {Bruderer}, S., {et~al.} 2012, \aap, 537,
  A55

\bibitem[{{Viti} \& {Williams}(1999)}]{Viti1999}
{Viti}, S., \& {Williams}, D.~A. 1999, \mnras, 310, 517

\bibitem[{{Ward-Thompson} {et~al.}(2017){Ward-Thompson}, {Pattle}, {Bastien},
  {Furuya}, {Kwon}, {Lai}, {Qiu}, {Berry}, {Choi}, {Coud{\'e}}, {Di Francesco},
  {Hoang}, {Franzmann}, {Friberg}, {Graves}, {Greaves}, {Houde}, {Johnstone},
  {Kirk}, {Koch}, {Kwon}, {Lee}, {Li}, {Matthews}, {Mottram}, {Parsons}, {Pon},
  {Rao}, {Rawlings}, {Shinnaga}, {Sadavoy}, {van Loo}, {Aso}, {Byun},
  {Eswaraiah}, {Chen}, {Chen}, {Chen}, {Ching}, {Cho}, {Chrysostomou}, {Chung},
  {Doi}, {Drabek-Maunder}, {Eyres}, {Fiege}, {Friesen}, {Fuller}, {Gledhill},
  {Griffin}, {Gu}, {Hasegawa}, {Hatchell}, {Hayashi}, {Holland}, {Inoue},
  {Inutsuka}, {Iwasaki}, {Jeong}, {Kang}, {Kang}, {Kang}, {Kawabata}, {Kemper},
  {Kim}, {Kim}, {Kim}, {Kim}, {Kim}, {Kim}, {Lacaille}, {Lee}, {Lee}, {Li},
  {Li}, {Liu}, {Liu}, {Liu}, {Liu}, {Lyo}, {Mairs}, {Matsumura},
  {Moriarty-Schieven}, {Nakamura}, {Nakanishi}, {Ohashi}, {Onaka}, {Peretto},
  {Pyo}, {Qian}, {Retter}, {Richer}, {Rigby}, {Robitaille}, {Savini}, {Scaife},
  {Soam}, {Tamura}, {Tang}, {Tomisaka}, {Wang}, {Wang}, {Whitworth}, {Yen},
  {Yoo}, {Yuan}, {Zhang}, {Zhang}, {Zhou}, {Zhu}, {Andr{\'e}}, {Dowell},
  {Falle}, \& {Tsukamoto}}]{WardThompson2017}
{Ward-Thompson}, D., {Pattle}, K., {Bastien}, P., {et~al.} 2017, \apj, 842, 66

\bibitem[{{Wootten}(1989)}]{Wootten1989}
{Wootten}, A. 1989, \apj, 337, 858

\bibitem[{Wurster \& Li(2018)}]{WursterLi2018}
Wurster, J., \& Li, Z.-Y. 2018, Frontiers in Astronomy and Space Sciences, 5,
  39

\bibitem[{{Yang} {et~al.}(2017){Yang}, {Evans}, {Green}, {Dunham}, \&
  {J{\o}rgensen}}]{YLYang2017}
{Yang}, Y.-L., {Evans}, II, N.~J., {Green}, J.~D., {Dunham}, M.~M., \&
  {J{\o}rgensen}, J.~K. 2017, \apj, 835, 259

\bibitem[{{Y{\i}ld{\i}z} {et~al.}(2015){Y{\i}ld{\i}z}, {Kristensen}, {van
  Dishoeck}, {Hogerheijde}, {Karska}, {Belloche}, {Endo}, {Frieswijk},
  {G{\"u}sten}, {van Kempen}, {Leurini}, {Nagy}, {P{\'e}rez-Beaupuits},
  {Risacher}, {van der Marel}, {van Weeren}, \& {Wyrowski}}]{Yildiz2015}
{Y{\i}ld{\i}z}, U.~A., {Kristensen}, L.~E., {van Dishoeck}, E.~F., {et~al.}
  2015, \aap, 576, A109

\bibitem[{Zhang {et~al.}(2014)Zhang, Qiu, Girart, Liu, Tang, Koch, Li, Keto,
  Ho, Rao, Lai, Ching, Frau, Chen, Li, Padovani, Bontemps, Csengeri, \&
  Ju{\'a}rez}]{Zhang2014}
Zhang, Q., Qiu, K., Girart, J.~M., {et~al.} 2014, The Astrophysical Journal,
  792, 116

\end{thebibliography}
\bibliographystyle{apj}

\end{document}